\documentclass[prd,aps,preprint,showpacs,preprintnumbers,amsmath,amssymb,tighten,nofootinbib,12pt]{revtex4}

\usepackage{graphicx}
\usepackage{dcolumn}
\usepackage{bm}
\usepackage{subfigure}

\allowdisplaybreaks

\begin{document}


\title{Non-unitary deviation from the tri-bimaximal lepton mixing and its
implications on neutrino oscillations \vspace{0.3cm}}

\author{\bf Shu Luo}
\email{luoshu@mail.ihep.ac.cn}


\affiliation{Institute of High Energy Physics, Chinese Academy of
Sciences, Beijing 100049, China \vspace{0.7cm}}

\begin{abstract}
We propose a new pattern of the neutrino mixing matrix which can
be parametrized as the product of an arbitrary Hermitian matrix
and the well-known tri-bimaximal mixing matrix. In this scenario,
nontrivial values of the smallest neutrino mixing angle
$\theta_{13}^{}$ and the CP-violating phases entirely arise from
the non-unitary corrections. We present a complete set of series
expansion formulas for neutrino oscillation probabilities both in
vacuum and in matter of constant density. We do a numerical
analysis to show the non-unitary effects on neutrino oscillations.
The possibility of determining small non-unitary perturbations and
CP-violating phases is discussed by measuring neutrino oscillation
probabilities and constructing ``deformed unitarity triangles".
Some brief comments on the non-unitary neutrino mixing matrix in
the type-II seesaw models are also given.
\end{abstract}

\pacs{11.30.Fs, 14.60.Pq, 14.60.St}

\maketitle

\section{Motivation}

Recent solar \cite{SNO}, atmospheric \cite{SK}, reactor \cite{KM}
and accelerator \cite{K2K} neutrino experiments have convincingly
verified the hypothesis of neutrino oscillations, which can
naturally happen if neutrinos are slightly massive and lepton
flavors are mixed. This discovery indicates the existence of new
physics beyond the Standard Model (SM). Although it is easy to add
either Dirac or Majorana neutrino mass terms to the SM, it is
highly non-trivial to reveal the essential meaning behind such
terms and find a natural and qualitative explanation of the
smallness of neutrino masses. A complete theory of neutrino mass
generation has been lacking and is eagerly desirable.

A low-energy effective theory responsible for the generation of
neutrino masses might give rise to slight violation of the
unitarity of the neutrino mixing matrix. If three light neutrinos
are mixed with other degrees of freedom (e.g., with the light
sterile neutrinos \cite{sterile}, the heavy Majorana neutrinos
\cite{seesaw}, or the whole tower of Kaluza-Klein states in some
models with extra dimensions \cite{extra dimension}), their
$3\times 3$ flavor mixing matrix $V$ appearing in the SM
charged-current interactions will in general be non-unitary.
Therefore, the deviation of $V$ from unitarity can serve as an
indicator of new physics beyond the SM.

A generic non-unitary neutrino mixing matrix $V$ can be parametrized
as $V = H \cdot V_{0}^{}$ \footnote{There is no unique way to
parametrize a generic $3 \times 3$ flavor mixing matrix. For
example, the non-unitary matrix $V$ can also be expressed as $V = A
\cdot V_{0}^{'}$, where $A$ is a lower (or upper) triangle matrix
and $V_{0}^{'}$ is a unitary matrix \cite{Xing}. Although different
parametrizations are mathematically equivalent, they may have their
own advantages in describing different phenomena of neutrino
physics.}, where $H$ is a Hermitian matrix which can be written as
\begin{equation}
H \; = \; \left ( \begin{matrix} a ~ & \hat{\kappa}_{12}^{} &
\hat{\kappa}_{13}^{} \cr \hat{\kappa}_{12}^{*} & b ~ &
\hat{\kappa}_{23}^{} \cr \hat{\kappa}_{13}^{*} &
\hat{\kappa}_{23}^{*} & c ~ \end{matrix} \right ) \; ,
\end{equation}
with $a, b, c$ being real and $\hat{\kappa}_{ij}^{}$ ($ij = 12,
13, 23$) being complex, and $V_{0}^{}$ is a unitary matrix which
is usually parametrized in terms of three mixing angles and one
Dirac CP-violating phase \cite{PDG} as
\begin{equation}
V_{0}^{} = \left( \begin{matrix} c_{12}^{} c_{13}^{} & s_{12}^{}
c_{13}^{} & s_{13}^{} e^{- i \delta} \cr - s_{12}^{} c_{23}^{} -
c_{12}^{} s_{23}^{} s_{13}^{} e^{i \delta} & c_{12}^{} c_{23}^{} -
s_{12}^{} s_{23}^{} s_{13}^{} e^{i \delta} & s_{23}^{} c_{13}^{} \cr
s_{12}^{} s_{23}^{} - c_{12}^{} c_{23}^{} s_{13}^{} e^{i \delta} & -
c_{12}^{} s_{23}^{} - s_{12}^{} c_{23}^{} s_{13}^{} e^{i \delta} &
c_{23}^{} c_{13}^{}
\end{matrix} \right)  \; .
\end{equation}
Here we have omitted two Majorana CP-violating phases of $V^{}_0$
since they are irrelevant to neutrino oscillations.

Given the lepton mixing matrix $V$, the constraints on the moduli
of the elements of $V V_{}^{\dagger}$ have been deduced in Ref.
\cite{Antusch} by combining the experimental data on both neutrino
oscillations and precision electroweak tests:
\begin{equation}
|V V_{}^{\dagger}| \approx \left ( \begin{matrix}
 0.994 \pm 0.005 & < 7.0 \times 10^{-5} & < 1.6 \times 10^{-2}
 \cr
 < 7.0 \times 10^{-5} & 0.995 \pm 0.005 & < 1.02 \times 10^{-2}
 \cr
 < 1.6 \times 10^{-2} & < 1.02 \times 10^{-2} & 0.995 \pm 0.005
\end{matrix} \right ) \, .
\end{equation}
It is then easy to find that $a, b, c \sim 1$,
$|\hat{\kappa}_{12}^{}| \lesssim 3.5 \times 10^{-5}$,
$|\hat{\kappa}_{13}^{}| \lesssim 8.0 \times 10^{-3}$, and
$|\hat{\kappa}_{23}^{}| \lesssim 5.1 \times 10^{-3}$ should hold.
We can further denote $a, b, c$ as $a = 1 + \epsilon_{a}^{}$, $b =
1 + \epsilon_{b}^{}$ and $c = 1 + \epsilon_{c}^{}$; namely,
\begin{equation} H \; \equiv \; \mathbf{1} + \mathbf{\epsilon} \;
= \; \mathbf{1} + \left (
\begin{matrix} \epsilon_{a}^{} & \hat{\kappa}_{12}^{} &
\hat{\kappa}_{13}^{} \cr \hat{\kappa}_{12}^{*} & \epsilon_{b}^{} &
\hat{\kappa}_{23}^{} \cr \hat{\kappa}_{13}^{*} &
\hat{\kappa}_{23}^{*} & \epsilon_{c}^{}
\end{matrix} \right ) \; ,
\end{equation}
where $\epsilon_{a}^{}$, $\epsilon_{b}^{}$, $\epsilon_{c}^{}$ are
all real. Bounds on $\epsilon_{a}^{}$, $\epsilon_{b}^{}$ and
$\epsilon_{c}^{}$ can also be obtained from Eq. (3):
$|\epsilon_{a}^{}| \lesssim 5.5 \times 10^{-3}$, $|\epsilon_{b}^{}|
\lesssim 5.0 \times 10^{-3}$ and $|\epsilon_{c}^{}| \lesssim 5.0
\times 10^{-3}$.

The effects of non-unitarity of $V$ on neutrino oscillations have
been discussed in some literature \cite{Xing,Antusch, non-unitarity,
Yasuda}. In particular, the authors of Ref. \cite{Yasuda} have used
the same parametrization of $V$ as given above and explored the
effects of CP violation induced by those non-unitary complex
parameters in neutrino oscillations.

In this paper, we start from an intriguing point of view that the
realistic neutrino mixing matrix $V$ might result from a
non-unitary correction to the well-known tri-bimaximal mixing
pattern \cite{tribi}. The latter is compatible with current
experimental data very well and can be derived from a number of
flavor symmetries and their spontaneous or explicit breaking
mechanisms \cite{deviationTB}. Instead of building a specific
neutrino model to realize such a phenomenological conjecture, here
we shall concentrate on the consequences of $V$ on neutrino
oscillations.

In our new neutrino mixing scenario, $V = H\cdot V^{}_0$ with
$V^{}_0$ being given by
\begin{equation}
V_{0}^{} \; = \; \left ( \begin{matrix} ~~ 2/\sqrt{6} & ~~
1/\sqrt{3} & ~ 0 ~ \cr - 1/\sqrt{6} & ~~ 1/\sqrt{3} & ~ 1/\sqrt{2}
~ \cr ~~ 1/\sqrt{6} & - 1/\sqrt{3} & ~ 1/\sqrt{2} ~ \end{matrix}
\right ) \;
\end{equation}
and $H$ being an arbitrary Hermitian matrix as shown in Eq. (1) or
(4). To be specific, the non-unitary neutrino mixing matrix $V$
reads
\begin{equation}
V \; = \; \left ( \begin{matrix} ~~ \displaystyle
\frac{2}{\sqrt{6}} ~ a - \frac{1}{\sqrt{6}} \left (
\hat{\kappa}_{12}^{} - \hat{\kappa}_{13}^{} \right ) &
\displaystyle ~~ \frac{1}{\sqrt{3}} ~ a + \frac{1}{\sqrt{3}} \left
( \hat{\kappa}_{12}^{} - \hat{\kappa}_{13}^{} \right ) & ~
\displaystyle \frac{1}{\sqrt{2}} \left ( \hat{\kappa}_{12}^{}
+\hat{\kappa}_{13}^{} \right ) ~ \cr \displaystyle -
\frac{1}{\sqrt{6}} ~ b + \frac{1}{\sqrt{6}} \left ( 2
\hat{\kappa}_{12}^{*} + \hat{\kappa}_{23}^{} \right ) & ~~
\displaystyle \frac{1}{\sqrt{3}} ~ b + \frac{1}{\sqrt{3}} \left (
\hat{\kappa}_{12}^{*} - \hat{\kappa}_{23}^{} \right ) & ~
\displaystyle \frac{1}{\sqrt{2}} ~ b + \frac{1}{\sqrt{2}}
\hat{\kappa}_{23}^{} ~ \cr ~~ \displaystyle \frac{1}{\sqrt{6}} ~ c
+ \frac{1}{\sqrt{6}} \left ( 2 \hat{\kappa}_{13}^{*} -
\hat{\kappa}_{23}^{*} \right ) & \displaystyle -
\frac{1}{\sqrt{3}} ~ c + \frac{1}{\sqrt{3}} \left (
\hat{\kappa}_{13}^{*} + \hat{\kappa}_{23}^{*} \right ) & ~
\displaystyle \frac{1}{\sqrt{2}} ~ c + \frac{1}{\sqrt{2}}
\hat{\kappa}_{23}^{*} ~ \end{matrix} \right ) \; .
\end{equation}
It is clear that the parameters of $H$ lead simultaneously to the
unitarity violation and the deviation from $V^{}_0$. The resulting
smallest mixing angle $\theta_{13}^{}$ to be measured in reactor
$\bar{\nu}_{e}^{} \rightarrow \bar{\nu}_{e}^{}$ oscillation
experiments is attributed to the small parameters
$\hat{\kappa}_{12}^{}$ and $\hat{\kappa}_{13}^{}$. So are the
CP-violating phases of $V$.

The remaining parts of this paper are organized as follows. In
sections II and III, we develop a complete set of series expansion
formulas for neutrino oscillation probabilities both in vacuum and
in matter of constant density, respectively, by taking account of
the non-unitary mixing matrix $V$ given in Eq. (6). In section IV,
we discuss the possibility of determining some parameters of $H$ by
constructing the ``deformed unitarity triangles". Section V is
devoted to a short discussion about incorporating our
parametrization of the non-unitary neutrino mixing matrix into a
generic type-II seesaw model. Finally some conclusions are drawn in
section VI.

\section{Neutrino oscillations in vacuum}

Suppose that the non-unitary V in Eq. (6) describes the mixing
between the neutrino fields in the mass basis and those in the
flavor basis,
\begin{equation}
\nu_{\alpha}^{} \; = \; V_{\alpha i}^{} ~ \nu_{i}^{} \; ,
\end{equation}
where $\alpha = e, \mu, \tau$ and $i = 1, 2, 3$. The probability
of neutrino oscillation $\nu_{\alpha}^{} \rightarrow
\nu_{\beta}^{}$ ($P_{\alpha \beta}^{}$) can be derived in a
similar way to that in the unitary case. The procedures of
deriving the formulas for neutrino oscillation probabilities can
be found in Ref. \cite{Antusch}. Here we follow another way, in
order to be concise. The derivation may be easily understood as
follows.

A typical neutrino oscillation process $\nu_{\alpha}^{} \rightarrow
\nu_{\beta}^{}$ can be divided into three parts \cite{Kayser}: 1)
$\nu_{\alpha}^{} $ being produced at the source through the
charged-current interaction which can be denoted as $W \rightarrow
\bar{l}_{\alpha}^{} \nu_{\alpha}^{}$. Here, $\nu_{\alpha}^{}$ is a
superposition of mass eigenstates $\nu_{i}^{}$; 2) $\nu_{i}^{}$
propagates from the source to the detector; 3) $\nu_{\beta}^{}$ (a
superposition of $\nu_{i}^{}$) being catched by the detector through
the charged-current interaction $\nu_{\beta}^{} W \rightarrow
l_{\beta}^{}$. Therefore, the amplitude of the neutrino oscillation
$\nu_{\alpha}^{} \rightarrow \nu_{\beta}^{}$ can be correspondingly
divided into three parts:
\begin{eqnarray}
A(\nu_{\alpha}^{} \rightarrow \nu_{\beta}^{}) = \sum_{i}^{} A(W
\rightarrow \bar{l}_{\alpha}^{} \nu_{i}^{}) {\rm Prop}(\nu_{i}^{})
A(\nu_{i}^{} W \rightarrow l_{\beta}^{}) \; .
\end{eqnarray}
In the case of non-unitary neutrino mixing, it follows that $A(W
\rightarrow \bar{l}_{\alpha}^{} \nu_{i}^{}) = V_{\alpha i}^{*} /
\sqrt{\left ( VV_{}^{\dagger} \right )_{\alpha \alpha}^{}}$. The
factor $1 / \sqrt{\left ( VV_{}^{\dagger} \right )_{\alpha
\alpha}^{}}$ ensures that the total rate $P(W \rightarrow
\bar{l}_{\alpha}^{} \nu_{\alpha}^{}) \equiv \sum_{i}^{} |A(W
\rightarrow \bar{l}_{\alpha}^{} \nu_{i}^{})|^2 = 1$. Similarly, we
have $A(\nu_{i}^{} W \rightarrow l_{\beta}^{}) = V_{\beta i}^{} /
\sqrt{\left ( VV_{}^{\dagger} \right )_{\beta \beta}^{}}$. The
expression of ${\rm Prop}(\nu_{i}^{})$ is the same as that in the
unitary case: ${\rm Prop}(\nu_{i}^{}) = \exp(-i m_{i}^{2} L /
2E_{\nu}^{})$. Finally, the amplitude of the neutrino oscillation
$\nu_{\alpha}^{} \rightarrow \nu_{\beta}^{}$ is given by
\begin{eqnarray}
A(\nu_{\alpha}^{} \rightarrow \nu_{\beta}^{}) = \frac{1}{\sqrt{\left
( V_{}^{} V_{}^{\dagger} \right )_{\alpha \alpha} \left ( V_{}^{}
V_{}^{\dagger} \right )_{\beta \beta}}} \sum_{i}^{} V_{\alpha i}^{*}
e_{}^{-i m_{i}^{2} \frac{L}{2E_{\nu}^{}}} V_{\beta i}^{} \; .
\end{eqnarray}
Then $P_{\alpha \beta}^{}$, the probability of neutrino
oscillation $\nu_{\alpha}^{} \rightarrow \nu_{\beta}^{}$, is given
by
\begin{eqnarray}
P_{\alpha \beta}^{} & \; = \; & |A(\nu_{\alpha}^{} \rightarrow
\nu_{\beta}^{})|^2 \; = \; \displaystyle \frac{\displaystyle \left|\sum_{i}^{}
V_{\alpha i}^{*} e_{}^{-i m_{i}^{2} \frac{L}{2E_{\nu}^{}}} V_{\beta
i}^{}\right|^2}{\left ( V_{}^{} V_{}^{\dagger} \right )_{\alpha
\alpha} \left ( V_{}^{} V_{}^{\dagger} \right )_{\beta \beta}} \; \nonumber\\
& \; = \; & \frac{1}{\left ( V_{}^{} V_{}^{\dagger} \right )_{\alpha
\alpha} \left ( V_{}^{} V_{}^{\dagger} \right )_{\beta \beta}} \left
[ |\left ( V_{}^{} V_{}^{\dagger} \right )_{\alpha \beta}|^2 - 4
\sum_{j<i} A_{\alpha \beta}^{ij} \sin^2 \Delta_{ij}^{} - 2
\sum_{j<i} J_{\alpha \beta}^{ij}
\sin2\Delta_{ij}^{} \right ] \; \nonumber\\
& \; = \; & \frac{1}{\left ( V_{}^{} V_{}^{\dagger} \right )_{\alpha
\alpha} \left ( V_{}^{} V_{}^{\dagger} \right )_{\beta \beta}} \cdot
\left [ |\left ( V_{}^{} V_{}^{\dagger} \right )_{\alpha \beta}|^2 -
4 A_{\alpha \beta}^{21} \sin^2 \Delta_{21}^{} - 4 A_{\alpha
\beta}^{31} \sin^2 \Delta_{31}^{} - 4 A_{\alpha
\beta}^{32} \sin^2 \Delta_{32}^{} \right. \nonumber\\
& & \hspace{4cm} \left . - 2 J_{\alpha \beta}^{21}
\sin2\Delta_{21}^{} - 2 J_{\alpha \beta}^{31} \sin2\Delta_{31}^{} -
2 J_{\alpha \beta}^{32} \sin2\Delta_{32}^{} \right ] \; ,
\end{eqnarray}
where $\Delta_{ij}^{} \equiv \Delta m_{ij}^{2}L/(4E_{\nu}^{})$
with $\Delta m_{ij}^{2} \equiv m_{i}^{2} - m_{j}^{2}$, and
$A_{\alpha \beta}^{ij} = {\rm Re} [ V_{\alpha i}^{} V_{\beta j}^{}
V_{\alpha j}^{*} V_{\beta i}^{*} ]$, $J_{\alpha \beta}^{ij} = {\rm
Im} [ V_{\alpha i}^{} V_{\beta j}^{} V_{\alpha j}^{*} V_{\beta
i}^{*} ]$. We can further absorb the renormalization factor $1 /
\left ( V_{}^{} V_{}^{\dagger} \right )_{\alpha \alpha} \left (
V_{}^{} V_{}^{\dagger} \right )_{\beta \beta}$ into the
redefinitions of $A_{\alpha \beta}^{ij}$ and $J_{\alpha
\beta}^{ij}$ and rewrite the above equation as
\begin{eqnarray}
P_{\alpha \beta}^{} & \; = \; & \frac{|\left ( V_{}^{}
V_{}^{\dagger} \right )_{\alpha \beta}|^2}{\left ( V_{}^{}
V_{}^{\dagger} \right )_{\alpha \alpha} \left ( V_{}^{}
V_{}^{\dagger} \right )_{\beta \beta}} - 4 \hat{A}_{\alpha
\beta}^{21} \sin^2 \Delta_{21}^{} - 4 \hat{A}_{\alpha \beta}^{31}
\sin^2 \Delta_{31}^{} - 4 \hat{A}_{\alpha \beta}^{32}
\sin^2 \Delta_{32}^{} \nonumber\\
& & - 2 \hat{J}_{\alpha \beta}^{21} \sin2\Delta_{21}^{} - 2
\hat{J}_{\alpha \beta}^{31} \sin2\Delta_{31}^{} - 2 \hat{J}_{\alpha
\beta}^{32} \sin2\Delta_{32}^{} \; ,
\end{eqnarray}
where $\hat{A}_{\alpha \beta}^{ij} = A_{\alpha \beta}^{ij} / (
V_{}^{} V_{}^{\dagger} )_{\alpha \alpha} ( V_{}^{} V_{}^{\dagger}
)_{\beta \beta}$, $\hat{J}_{\alpha \beta}^{ij} = J_{\alpha
\beta}^{ij} / ( V_{}^{} V_{}^{\dagger} )_{\alpha \alpha} ( V_{}^{}
V_{}^{\dagger} )_{\beta \beta}$.

The first term in Eq. (11) is the so-called ``zero-distance" term.
It means that at $L = 0$ we have
\begin{equation}
P_{\alpha \beta}^{} (L = 0) \; = \; \frac{|\left ( V_{}^{}
V_{}^{\dagger} \right )_{\alpha \beta}|^2}{\left ( V_{}^{}
V_{}^{\dagger} \right )_{\alpha \alpha} \left ( V_{}^{}
V_{}^{\dagger} \right )_{\beta \beta}}  \; .
\end{equation}
Note that if $\alpha = \beta$, $P_{\alpha \beta}^{} (L = 0) = 1$;
namely, there are no ``zero-distance" effects in the disappearance
experiments. If $\alpha \neq \beta$,  the oscillation probability
$P_{\alpha \beta}^{} (L = 0)$ is in general nonzero, that is the
``zero-distance" effect. One can find that although it is nonzero,
this term add only a tiny constant to the oscillation probability,
and does not change the oscillatory behavior. In our scenario, we
have $P_{e \mu}^{} (L = 0) \approx 4 |\hat{\kappa}_{12}^{}|^2$,
$P_{e \tau}^{} (L = 0) \approx 4 |\hat{\kappa}_{13}^{}|^2$ and
$P_{\mu \tau}^{} (L = 0) \approx 4 |\hat{\kappa}_{23}^{}|^2$.
Another significant difference between the non-unitary and unitary
cases is: if the mixing matrix $V$ is non-unitary, there may exist
9 different Jarlskog invariants $J_{\alpha \beta}^{ij}$
corresponding to 3 different oscillation channels instead of a
unique Jarlskog in the unitary case.

If $\Delta_{21}^{} \ll 1$ is satisfied, we can expand Eq. (11) as
\begin{eqnarray}
P_{\alpha \beta}^{} & \; \approx \; & \frac{|\left ( V_{}^{}
V_{}^{\dagger} \right )_{\alpha \beta}|^2}{\left ( V_{}^{}
V_{}^{\dagger} \right )_{\alpha \alpha} \left ( V_{}^{}
V_{}^{\dagger} \right )_{\beta \beta}} \nonumber\\
& & - 4 \left ( \hat{A}_{\alpha \beta}^{21} + \hat{A}_{\alpha
\beta}^{32} \right ) \Delta_{21}^{2} - 4 \left ( \hat{A}_{\alpha
\beta}^{31} + \hat{A}_{\alpha \beta}^{32} \right ) \sin^2
\Delta_{31}^{} \nonumber \\
& & + 4 \hat{A}_{\alpha \beta}^{32} \left ( \Delta_{21}^{} \sin 2
\Delta_{31}^{} + 2 \Delta_{21}^{2}
\sin^2 \Delta_{31}^{} \right ) \nonumber\\
& & - 4 \left ( \hat{J}_{\alpha \beta}^{21} - \hat{J}_{\alpha
\beta}^{32} \right ) \Delta_{21}^{} - 2 \left ( \hat{J}_{\alpha
\beta}^{31} + \hat{J}_{\alpha \beta}^{32} \right )
\sin2\Delta_{31}^{} \nonumber \\
& & - 4 \hat{J}_{\alpha \beta}^{32} \left ( \Delta_{21}^{2} \sin 2
\Delta_{31}^{} - 2 \Delta_{21}^{} \sin^2 \Delta_{31}^{} \right ) \;
, ~~~~\;
\end{eqnarray}
In our calculations, we find that although all the nine
$\hat{J}_{\alpha \beta}^{ij}$ are of ${\cal
O}(\hat{\kappa}_{ij}^{})$, only $\hat{J}_{\mu \tau}^{32} +
\hat{J}_{\mu \tau}^{31}$ is of ${\cal O}(\hat{\kappa}_{ij}^{})$
while $\hat{J}_{e \mu}^{32} + \hat{J}_{e \mu}^{31}$ and $\hat{J}_{e
\tau}^{32} + \hat{J}_{e \tau}^{31}$ are both of ${\cal
O}(\hat{\kappa}_{ij}^{2})$. This observation means that the most
sensitive way at short baseline neutrino oscillation experiments to
detect CP violation is to measure the $\nu_{\mu}^{} \rightarrow
\nu_{\tau}^{}$ channel. Such a point was also pointed out in Refs.
\cite{Xing,Yasuda}.

Here we present a complete set of formulas for neutrino oscillation
probabilities $P_{\alpha \beta}^{}$ to the second order in powers of
$\hat{\kappa}_{12}^{}$, $\hat{\kappa}_{13}^{}$,
$\hat{\kappa}_{23}^{}$, $\epsilon_{a}^{}$, $\epsilon_{b}^{}$,
$\epsilon_{c}^{}$ and $\Delta_{21}^{}$. These formulas are good
approximations for the $\Delta_{31}^{}$-dominated oscillations,
i.e., for neutrino oscillation experiments with relative short
baselines and relatively high energies.
\begin{eqnarray}
P_{ee}^{} & \; \approx \; & 1 - 2 |\hat{\kappa}_{12}^{} +
\hat{\kappa}_{13}^{}|_{}^{2} \sin^2 \Delta_{31}^{} - \frac{8}{9}
\Delta_{21}^{2} \; ,\\
P_{\mu \mu}^{} & \; \approx \; & 1 - \left [ 1 - 4 \left ( {\rm
Re}[\hat{\kappa}_{23}^{}] \right )^2 \right ] \sin^2 \Delta_{31}^{}
+ \frac{2}{3} \left ( 1 + 2 {\rm Re}[\hat{\kappa}_{12}^{}] \right )
\Delta_{21}^{} \sin2\Delta_{31}^{} \nonumber\\
& & - \frac{4}{9} \left ( 2 - 3 \sin^2 \Delta_{31}^{} \right )
\Delta_{21}^{2} \; ,
\\
P_{\tau \tau}^{} & \; \approx \; & 1 - \left ( 1 - 4 \left ( {\rm
Re}[\hat{\kappa}_{23}^{}] \right )^2 \right ) \sin^2 \Delta_{31}^{}
+ \frac{2}{3} \left ( 1 - 2 {\rm Re}[\hat{\kappa}_{13}^{}] \right )
\Delta_{21}^{} \sin2\Delta_{31}^{} \nonumber\\
& & - \frac{4}{9} \left ( 2 - 3 \sin^2 \Delta_{31}^{} \right )
\Delta_{21}^{2} \; ,
\\
P_{e \mu}^{} & \; \approx \; & 4 |\hat{\kappa}_{12}^{}|^2 - \left (
3 |\hat{\kappa}_{12}^{}|^2 - |\hat{\kappa}_{13}^{}|^2 + 2 {\rm
Re}[\hat{\kappa}_{12}^{} \hat{\kappa}_{13}^{*}] \right ) \sin^2
\Delta_{31}^{} + \frac{4}{9} \Delta_{21}^{2} \nonumber\\
& & + \frac{2}{3} {\rm Re}[\hat{\kappa}_{12}^{} +
\hat{\kappa}_{13}^{}] \Delta_{21}^{} \sin2\Delta_{31}^{} + 2 {\rm
Im}[\hat{\kappa}_{12}^{} \hat{\kappa}_{13}^{*}]
\sin2\Delta_{31}^{} \nonumber\\
& & + \frac{4}{3} \left ( 2 {\rm Im}[\hat{\kappa}_{12}^{}] - {\rm
Im}[\hat{\kappa}_{12}^{} + \hat{\kappa}_{13}^{}] \sin^2
\Delta_{31}^{} \right ) \Delta_{21}^{} \; ,\\
P_{e \tau}^{} & \; \approx \; & 4 |\hat{\kappa}_{13}^{}|^2 - \left (
3 |\hat{\kappa}_{13}^{}|^2 - |\hat{\kappa}_{12}^{}|^2 + 2 {\rm
Re}[\hat{\kappa}_{12}^{} \hat{\kappa}_{13}^{*}] \right ) \sin^2
\Delta_{31}^{} + \frac{4}{9} \Delta_{21}^{2} \nonumber\\
& & - \frac{2}{3} {\rm Re}[\hat{\kappa}_{12}^{} +
\hat{\kappa}_{13}^{}] \Delta_{21}^{} \sin2\Delta_{31}^{} - 2 {\rm
Im}[\hat{\kappa}_{12}^{} \hat{\kappa}_{13}^{*}]
\sin2\Delta_{31}^{} \nonumber\\
& & - \frac{4}{3} \left ( 2 {\rm Im}[\hat{\kappa}_{13}^{}] - {\rm
Im}[\hat{\kappa}_{12}^{} + \hat{\kappa}_{13}^{}] \sin^2
\Delta_{31}^{} \right ) \Delta_{21}^{} \; ,\\
P_{\mu \tau}^{} & \; \approx \; & 4 |\hat{\kappa}_{23}^{}|^2 + \left
[ 1 - |\hat{\kappa}_{12}^{} + \hat{\kappa}_{13}^{}|^2 - 4 \left (
|\hat{\kappa}_{23}^{}|^2 + \left ( {\rm Im}[\hat{\kappa}_{23}^{}]
\right )^2 \right
) \right ] \sin^2 \Delta_{31}^{} \nonumber\\
& & - \frac{2}{3} \left ( 1 + {\rm Re}[\hat{\kappa}_{12}^{} -
\hat{\kappa}_{13}^{}] \right ) \Delta_{21}^{}
\sin2\Delta_{31}^{} + \frac{4}{9} \left ( 1 - 3 \sin^2 \Delta_{31}^{} \right ) \Delta_{21}^{2} \nonumber\\
& & + \left [ \left ( 2 - \epsilon_{b}^{} - \epsilon_{c}^{} \right )
{\rm Im}[\hat{\kappa}_{23}^{}] - {\rm Im}[\hat{\kappa}_{12}^{}
\hat{\kappa}_{13}^{*}] \right ] \sin2\Delta_{31}^{} \nonumber\\
& & - \frac{4}{3} \left ( 2 {\rm Im}[\hat{\kappa}_{23}^{}] - {\rm
Im}[\hat{\kappa}_{12}^{} + \hat{\kappa}_{13}^{} +
\hat{\kappa}_{23}^{}] \sin^2 \Delta_{31}^{} \right )\Delta_{21}^{}
\; .
\end{eqnarray}
The first term in each of the above six equations is the
``zero-distance" term. The last two terms in Eq. (17), (18) or
(19) are the ``CP-violating" terms.

Suppose that the absolute values of those non-unitary parameters are
around their upper bounds, i.e., $|\hat{\kappa}_{12}^{}| \sim 3.5
\times 10^{-5}$, $|\hat{\kappa}_{13}^{}| \sim 8.0 \times 10^{-3}$,
$|\hat{\kappa}_{23}^{}| \sim 5.0 \times 10^{-3}$, $|\epsilon_{a}^{}|
\sim 5.5 \times 10^{-3}$, $|\epsilon_{b}^{}| \sim 5.0 \times
10^{-3}$ and $|\epsilon_{c}^{}| \sim 5.0 \times 10^{-3}$. We notice
that $P_{ee}^{}$ is only sensitive to $|V_{e3}^{}| =
|\hat{\kappa}_{12}^{} + \hat{\kappa}_{13}^{}| / \sqrt{2}$, therefore
we are able to determine $|\hat{\kappa}_{13}^{}|$ through the
detection of $\nu_{e}^{} \rightarrow \nu_{e}^{}$ oscillation. By
measuring the probability of $\nu_{\mu}^{} \rightarrow \nu_{\mu}^{}$
oscillation it is possible to determine or constrain the value of
${\rm Re}[\hat{\kappa}_{23}^{}]$. If the small difference between
$P_{\tau \tau}^{}$ and $P_{\mu \mu}^{}$ can be well measured, then
we are able to obtain the information on ${\rm
Re}[\hat{\kappa}_{13}^{}]$. Combined with the value of
$|\hat{\kappa}_{13}^{}|$, ${\rm arg}(\hat{\kappa}_{13}^{})$ can be
determined. As for ${\rm Im}[\hat{\kappa}_{23}^{}]$, the most
effective way is to probe the CP-violating terms in the
$\nu_{\mu}^{} \rightarrow \nu_{\tau}^{}$ channel.

\section{Neutrino oscillations in matter}

When a neutrino beam passes through matter, only $\nu^{}_e$ can
interact with electrons in the medium via the charged-current
interactions, while $\nu^{}_e$, $\nu^{}_\mu$ and $\nu^{}_\tau$ can
all interact with electrons, protons and neutrons in the medium via
the neutral-current interactions. The coherent forward scattering
from the constituents of matter modifies the evolution behaviors of
the neutrino beam. In the vacuum mass eigenbasis, the evolution
equation of neutrinos can be written as
\begin{equation}
i{\frac{d}{dt}} | \nu_{m}^{}(t)\rangle = \tilde{{\cal H}} |
\nu_{m}^{}(t)\rangle \; .
\end{equation}
We use tildes to denote the quantities in matter. For the
propagation of neutrinos in matter of constant density, the
Hamiltonian $\tilde{{\cal H}}$ is given by \cite{MSW}
\begin{equation}
\tilde{{\cal H}} = E + V_{}^{T} A V_{}^{*} \; ,
\end{equation}
where $E \equiv {\rm diag}(E_{1}^{}, E_{2}^{}, E_{3}^{})$ is the
energy matrix in the mass eigenbasis in vacuum, $A \equiv {\rm
diag}(V_{CC}^{} - V_{NC}^{}, - V_{NC}^{}, - V_{NC}^{})$, with
$V_{CC}^{} \equiv \sqrt{2} G_{F}^{} n_{e}^{}$ and $V_{NC}^{} \equiv
\displaystyle \frac{1}{\sqrt{2}} G_{F}^{} n_{n}^{}$ ($n_{e}^{}$ and
$n_{n}^{}$ are the electron and neutron densities, respectively).
Here $V$ is just the non-unitarity mixing matrix in Eq. (6).

The Hermitian matrix $\tilde{{\cal H}}$ can be diagonalized by a
unitary transformation $\tilde{{\cal H}} = U_{}^{} \tilde{E}
U_{}^{\dagger}$, where $\tilde{E} \equiv {\rm
diag}(\tilde{E}_{1}^{}, \tilde{E}_{2}^{}, \tilde{E}_{3}^{})$ is
the effective energy matrix in matter. The solution to Eq. (20)
can be expressed as
\begin{equation}
| \nu_{m}^{}(L)\rangle = U e_{}^{-i \tilde{E} L} U_{}^{\dagger} |
\nu_{m}^{}(0)\rangle \; ,
\end{equation}
where we have inserted $L=t$. From Eq. (22) we can work out the
neutrino oscillation probabilities in matter:
\begin{eqnarray}
\tilde{P}_{\alpha \beta}^{} & \; = \; & \frac{\left | \left (
V_{}^{*} U_{}^{} e_{}^{-i \tilde{E}_{}^{} L} U_{}^{\dagger} V_{}^{T}
\right )_{\alpha \beta} \right |^2}{\left ( V_{}^{} V_{}^{\dagger}
\right )_{\alpha \alpha} \left ( V_{}^{} V_{}^{\dagger} \right
)_{\beta \beta}} \; = \; \frac{\left | \left ( X_{}^{*} e_{}^{-i
\tilde{E}_{}^{} L} X_{}^{T} \right )_{\alpha \beta} \right
|^2}{\left ( X_{}^{} X_{}^{\dagger} \right )_{\alpha
\alpha} \left ( X_{}^{} X_{}^{\dagger} \right )_{\beta \beta}} \; \nonumber\\
& \; = \; & \frac{1}{\left ( X_{}^{} X_{}^{\dagger} \right )_{\alpha
\alpha} \left ( X_{}^{} X_{}^{\dagger} \right )_{\beta \beta}} \left
[ |\left ( X_{}^{} X_{}^{\dagger} \right )_{\alpha \beta}|^2 - 4
\tilde{A}_{\alpha \beta}^{21} \sin^2 \tilde{\Delta}_{21}^{} - 4
\tilde{A}_{\alpha \beta}^{31} \sin^2 \tilde{\Delta}_{31}^{} - 4
\tilde{A}_{\alpha \beta}^{32} \sin^2 \tilde{\Delta}_{32}^{} \right. \nonumber\\
& & \hspace{4cm} \left . - 2 \tilde{J}_{\alpha \beta}^{21}
\sin2\tilde{\Delta}_{21}^{} - 2 \tilde{J}_{\alpha \beta}^{31}
\sin2\tilde{\Delta}_{31}^{} - 2 \tilde{J}_{\alpha \beta}^{32}
\sin2\tilde{\Delta}_{32}^{} \right ] \; ,
\end{eqnarray}
where $X \equiv V U_{}^{*}$, $\tilde{A}_{\alpha \beta}^{ij} = {\rm
Re} [ X_{\alpha i}^{} X_{\beta j}^{} X_{\alpha j}^{*} X_{\beta
i}^{*} ]$, $\tilde{J}_{\alpha \beta}^{ij} = {\rm Im} [ X_{\alpha
i}^{} X_{\beta j}^{} X_{\alpha j}^{*} X_{\beta i}^{*} ]$ and
$\tilde{\Delta}_{ij}^{} \equiv \displaystyle
\frac{\tilde{E}_{i}^{} - \tilde{E}_{j}^{}}{2}$. Comparing Eq. (23)
with Eq. (11), we find that the matrix $X$, which is also
non-unitary, can be regarded as the effective neutrino mixing
matrix in matter.

In Appendix A, we present the details of the approximate
diagonalization of the Hamiltonian $\tilde{{\cal H}}$ by using the
perturbation theory. In the results to be presented below, those
terms in proportion to $\left ( V_{CC}^{} - 2V_{NC}^{} \right )$
will be neglected. The reason is simple: for ordinary earth matter,
which is electrically neutral, we have $n_{e}^{} \approx n_{n}^{}$
to a good degree of accuracy, and thus we can safely set $V_{CC}^{}
- 2 V_{NC}^{} = \sqrt{2} G_{F}^{} \left ( n_{e}^{} - n_{n}^{} \right
) \approx 0$. It is necessary to mention that the subsequent
analytical approximations are not very good for a relative large $L
/ E_{\nu}^{}$ or for the $\Delta m_{21}^{2}$-dominated oscillation.
In addition, we cannot directly obtain Eqs. (14) $\sim$ (19) from
Eqs. (27) $\sim$ (32) by setting $V_{CC}^{}, V_{NC}^{} \rightarrow
0$. This is because the expansion of $\tilde{\cal{H}}$ in Eqs. (A4)
$\sim$ (A7) is improper if $V_{CC}^{} = V_{NC}^{} = 0$.

The eigenvalues of $\tilde{\cal H}$, $\tilde{E}_{i}^{}$ $(i = 1, 2,
3)$, are related to the effective neutrino masses in matter by the
relations $\tilde{E}_{i}^{} \approx E_{\nu}^{} + \displaystyle
\frac{\tilde{\lambda}_{i}^{2}}{2 E_{\nu}^{}}$, where $E_{\nu}^{}$ is
the energy of neutrinos \footnote{Here we use the notation
$\tilde{\lambda}_{i}^{}$ and $\Delta \tilde{\lambda}_{ij}^{2}$ to
denote the effective neutrino masses and the mass squared
differences in matter instead of $\tilde{m}_{i}^{}$ and $\Delta
\tilde{m}_{ij}^{2}$, because we did not ordering $\tilde{E}_{i}^{}$
according to their magnitude and the mass spectrum. After ordering
$\tilde{E}_{i}^{}$, we have $\tilde{\lambda}_{i}^{} =
\tilde{m}_{i}^{}$ and $\Delta \tilde{\lambda}_{ij}^{2} = \Delta
\tilde{m}_{ij}^{2}$.}. Then the effective mass squared differences
in matter are given by $\Delta \tilde{\lambda}_{ij}^{2} \equiv 2
E_{\nu}^{} ( \tilde{E}_{i}^{} - \tilde{E}_{j}^{} )$ which are shown
in Eqs. (24), (25) and (26) to the second order in
$\hat{\kappa}_{12}^{}$, $\hat{\kappa}_{13}^{}$,
$\hat{\kappa}_{23}^{}$, $\epsilon_{a}^{}$, $\epsilon_{b}^{}$,
$\epsilon_{c}^{}$ and $\Delta_{21}^{}$.
\begin{eqnarray}
\Delta \tilde{\lambda}_{21}^{2} & \; \approx \; & - 2 E_{\nu}^{}
V_{CC}^{} + \frac{1}{3} \Delta m_{21}^{2} - 4 E_{\nu}^{} \left (
V_{CC}^{} - V_{NC}^{} \right ) \epsilon_{a}^{} \nonumber\\
& & - 2 E_{\nu}^{} V_{NC}^{} \left ( \epsilon_{b}^{} +
\epsilon_{c}^{} + 2 {\rm Re} [\hat{\kappa}_{23}^{}] \right ) -
\frac{2 \left ( \Delta m_{21}^{2} \right )^2}{9 E_{\nu}^{}
V_{CC}^{}}
\nonumber\\
& & - \frac{\left ( 2 E_{\nu}^{} V_{NC}^{} \right )^2}{\Delta
m_{31}^{2}} \left [ \left ( \epsilon_{b}^{} - \epsilon_{c}^{} \right
)^2 + 4 \left ( {\rm Im}
[\hat{\kappa}_{23}^{}] \right )^2 \right ] \nonumber\\
& & + E_{\nu}^{} \left ( V_{CC}^{} - V_{NC}^{} \right ) \left (
\left | \hat{\kappa}_{12}^{} - \hat{\kappa}_{13}^{} \right |^2 - 2
\epsilon_{a}^{2} \right ) \nonumber\\
& & - E_{\nu}^{} V_{NC}^{} \left ( \left | \epsilon_{b}^{} -
\hat{\kappa}_{23}^{} \right |^2 + \left | \epsilon_{c}^{} -
\hat{\kappa}_{23}^{} \right |^2  - 2 \left | \hat{\kappa}_{12}^{}
\right |^2 - 2 \left |
\hat{\kappa}_{13}^{} \right |^2 \right ) \; , \\
\Delta \tilde{\lambda}_{31}^{2} & \; \approx \; & \Delta m_{31}^{2}
- 2 E_{\nu}^{} V_{CC}^{} - \frac{1}{3} \Delta m_{21}^{2} - 4
E_{\nu}^{} \left ( V_{CC}^{} - V_{NC}^{} \right ) \epsilon_{a}^{}
\nonumber\\
& & - 2 E_{\nu}^{} V_{NC}^{} \left ( \epsilon_{b}^{} +
\epsilon_{c}^{} - 2 {\rm Re} [\hat{\kappa}_{23}^{}] \right ) -
\frac{\left ( \Delta m_{21}^{2}
\right )^2}{9 E_{\nu}^{} V_{CC}^{}} \nonumber\\
& & + \frac{\left ( 2 E_{\nu}^{} V_{NC}^{} \right )^2}{\Delta
m_{31}^{2}} \left [ \left ( \epsilon_{b}^{} - \epsilon_{c}^{} \right
)^2 + 4 \left ( {\rm Im}
[\hat{\kappa}_{23}^{}] \right )^2 \right ] \nonumber\\
& & + E_{\nu}^{} \left ( V_{CC}^{} - V_{NC}^{} \right ) \left (
\left | \hat{\kappa}_{12}^{} + \hat{\kappa}_{13}^{} \right |^2 - 2
\epsilon_{a}^{2} \right ) \nonumber\\
& & - E_{\nu}^{} V_{NC}^{} \left ( \left | \epsilon_{b}^{} +
\hat{\kappa}_{23}^{} \right |^2 + \left | \epsilon_{c}^{} +
\hat{\kappa}_{23}^{} \right |^2  - 2 \left | \hat{\kappa}_{12}^{}
\right |^2 - 2 \left |
\hat{\kappa}_{13}^{} \right |^2 \right ) \; , \\
\Delta \tilde{\lambda}_{32}^{2} & \; \approx \; & \Delta m_{31}^{2}
- \frac{2}{3} \Delta m_{21}^{2} + 8 E_{\nu}^{} V_{NC}^{} {\rm Re}
[\hat{\kappa}_{23}^{}] + \frac{\left ( \Delta m_{21}^{2} \right
)^2}{9 E_{\nu}^{}
V_{CC}^{}} \nonumber\\
& & + \frac{2 \left ( 2 E_{\nu}^{} V_{NC}^{} \right )^2}{\Delta
m_{31}^{2}} \left [ \left ( \epsilon_{b}^{} - \epsilon_{c}^{} \right
)^2 + 4 \left ( {\rm Im}
[\hat{\kappa}_{23}^{}] \right )^2 \right ] \nonumber\\
& & + 4 E_{\nu}^{} \left ( V_{CC}^{} - V_{NC}^{} \right ) {\rm Re}[
\hat{\kappa}_{12}^{} \hat{\kappa}_{13}^{*}] - 4 E_{\nu}^{} V_{NC}^{}
\left ( \epsilon_{b}^{} + \epsilon_{c}^{} \right ) {\rm Re}[
\hat{\kappa}_{23}^{}] \; .
\end{eqnarray}

In Eqs. (27) to (32), we present the expansion forms of Eq. (22)
for all six neutrino oscillation probabilities to the second order
in $\hat{\kappa}_{12}^{}$, $\hat{\kappa}_{13}^{}$,
$\hat{\kappa}_{23}^{}$, $\epsilon_{a}^{}$, $\epsilon_{b}^{}$,
$\epsilon_{c}^{}$ and $\Delta_{21}^{}$ in terms of the quantities
in vacuum.
\begin{eqnarray}
\tilde{P}_{ee}^{} & \; = \; & 1 - 2 \left | \hat{\kappa}_{12}^{} +
\hat{\kappa}_{13}^{} \right |^2 \sin^2 \left ( \Delta_{31}^{} -
\frac{V_{CC}^{} L}{2} \right ) \nonumber\\
& & - 2 \left | \frac{4 \Delta_{21}^{}}{3 V_{CC}^{} L} - \left (
\hat{\kappa}_{12}^{} - \hat{\kappa}_{13}^{} \right ) \right |^2
\sin^2 \left ( \frac{V_{CC}^{} L}{2} \right ) \; ,~~~~~~ \\
\nonumber\\
\tilde{P}_{\mu \mu}^{} & \; = \; & 1 - \left [ 1 - 8 \left (
\frac{1}{3} \Delta_{21}^{} + 2 V_{NC}^{} L {\rm
Re}[\hat{\kappa}_{23}^{}] \right )^2 + \frac{2 \sqrt{2}
\Delta_{21}^{} V_{NC}^{} L}{3 \Delta_{31}^{} \left ( V_{CC}^{} - 2
\Delta_{31}^{} \right )} \left ( \epsilon_{b}^{} -\epsilon_{c}^{}
\right ) \right. \nonumber\\
& & \left. - 4 \left ( {\rm Re}[\hat{\kappa}_{23}^{}] \right )^2 +
\frac{4 V_{CC}^{} L}{\Delta_{31}^{}} \left ( \epsilon_{b}^{}
-\epsilon_{c}^{} \right ) {\rm Re}[\hat{\kappa}_{23}^{}] -
\frac{\left ( V_{CC}^{} L \right )^2}{\Delta_{31}^{}} \left (
\epsilon_{b}^{} -\epsilon_{c}^{} \right )^2 \right ] \sin^2
\Delta_{31}^{}
\nonumber\\
& & + \left [ 2 \left ( \frac{1}{3} \Delta_{21}^{} + 2 V_{NC}^{} L
{\rm Re}[\hat{\kappa}_{23}^{}] \right ) - \frac{4 \Delta_{21}^{2}}{9
V_{CC}^{} L} - 2 \left ( V_{CC}^{} - V_{NC}^{} \right ) L {\rm
Re}[\hat{\kappa}_{12}^{} \hat{\kappa}_{13}^{*}] \right. \nonumber\\
& & \left. + 2 V_{NC}^{} L \left ( \epsilon_{b}^{} + \epsilon_{c}^{}
\right ) {\rm Re}[\hat{\kappa}_{23}^{}] - \frac{\left ( V_{NC}^{} L
\right )^2}{\Delta_{31}^{}} \left [ \left ( \epsilon_{b}^{} -
\epsilon_{c}^{} \right )^2 + 4 \left ( {\rm Im}
[\hat{\kappa}_{23}^{}] \right )^2 \right ] \right ]
\sin2\Delta_{31}^{}
\nonumber\\
& & + 2 \left | \frac{2 \Delta_{21}^{}}{3 V_{CC}^{} L} +
\hat{\kappa}_{12}^{} \right |^2 \cdot \left [ 2 \sin^2
\Delta_{31}^{} \sin^2 \left ( \frac{V_{CC}^{} L}{2} \right ) +
\frac{1}{2} \sin2 \Delta_{31}^{} \sin \left ( V_{CC}^{} L \right )
\right ]\nonumber\\
& & - 4 \left | \frac{2 \Delta_{21}^{}}{3 V_{CC}^{} L} +
\hat{\kappa}_{12}^{} \right |^2 \sin^2 \left ( \frac{V_{CC}^{} L}{2}
\right ) - 4 \left ( \frac{1}{3} \Delta_{21}^{} + 2 V_{NC}^{} L {\rm
Re}[\hat{\kappa}_{23}^{}] \right )^2 \; ,\\
\nonumber\\
\tilde{P}_{\tau \tau}^{} & \; = \; & 1 - \left [ 1 - 8 \left (
\frac{1}{3} \Delta_{21}^{} + 2 V_{NC}^{} L {\rm
Re}[\hat{\kappa}_{23}^{}] \right )^2 - \frac{2 \sqrt{2}
\Delta_{21}^{} V_{NC}^{} L}{3 \Delta_{31}^{} \left ( V_{CC}^{} - 2
\Delta_{31}^{} \right )} \left ( \epsilon_{b}^{} -\epsilon_{c}^{}
\right ) \right. \nonumber\\
& & \left. - 4 \left ( {\rm Re}[\hat{\kappa}_{23}^{}] \right )^2 -
\frac{4 V_{CC}^{} L}{\Delta_{31}^{}} \left ( \epsilon_{b}^{}
-\epsilon_{c}^{} \right ) {\rm Re}[\hat{\kappa}_{23}^{}] -
\frac{\left ( V_{CC}^{} L \right )^2}{\Delta_{31}^{}} \left (
\epsilon_{b}^{} -\epsilon_{c}^{} \right )^2 \right ] \sin^2
\Delta_{31}^{}
\nonumber\\
& & + \left [ 2 \left ( \frac{1}{3} \Delta_{21}^{} + 2 V_{NC}^{} L
{\rm Re}[\hat{\kappa}_{23}^{}] \right ) - \frac{4 \Delta_{21}^{2}}{9
V_{CC}^{} L} - 2 \left ( V_{CC}^{} - V_{NC}^{} \right ) L {\rm
Re}[\hat{\kappa}_{12}^{} \hat{\kappa}_{13}^{*}] \right. \nonumber\\
& & \left. + 2 V_{NC}^{} L \left ( \epsilon_{b}^{} + \epsilon_{c}^{}
\right ) {\rm Re}[\hat{\kappa}_{23}^{}] - \frac{\left ( V_{CC}^{} L
\right )^2}{\Delta_{31}^{}} \left [ \left ( \epsilon_{b}^{} -
\epsilon_{c}^{} \right )^2 + 4 \left ( {\rm Im}
[\hat{\kappa}_{23}^{}] \right )^2 \right ] \right ]
\sin2\Delta_{31}^{}
\nonumber\\
& & + 2 \left | \frac{2 \Delta_{21}^{}}{3 V_{CC}^{} L} -
\hat{\kappa}_{13}^{} \right |^2 \cdot \left [ 2 \sin^2
\Delta_{31}^{} \sin^2 \left ( \frac{V_{CC}^{} L}{2} \right ) +
\frac{1}{2} \sin2 \Delta_{31}^{} \sin \left ( V_{CC}^{} L \right )
\right ]\nonumber\\
& & - 4 \left | \frac{2 \Delta_{21}^{}}{3 V_{CC}^{} L} -
\hat{\kappa}_{13}^{} \right |^2 \sin^2 \left ( \frac{V_{CC}^{} L}{2}
\right ) - 4 \left ( \frac{1}{3} \Delta_{21}^{} + 2 V_{NC}^{} L {\rm
Re}[\hat{\kappa}_{23}^{}] \right )^2 \; , \\
\nonumber\\
\tilde{P}_{e \mu}^{} & \; = \; & 4 |\hat{\kappa}_{12}^{}|^2 - \left
( 3 |\hat{\kappa}_{12}^{}|^2 - |\hat{\kappa}_{13}^{}|^2 + 2 {\rm
Re}[\hat{\kappa}_{12}^{} \hat{\kappa}_{13}^{*}] \right )
\sin^2 \Delta_{31}^{} \nonumber\\
& & + 4 \left ( \frac{2 \Delta_{21}^{}}{3 V_{CC}^{} L} + {\rm
Re}[\hat{\kappa}_{12}^{}] \right ) {\rm Re}[\hat{\kappa}_{12}^{} +
\hat{\kappa}_{13}^{}] \sin^2 \Delta_{31}^{} \sin^2 \left (
\frac{V_{CC}^{} L}{2} \right ) \nonumber\\
& & + \left ( \frac{2 \Delta_{21}^{}}{3 V_{CC}^{} L} + {\rm
Re}[\hat{\kappa}_{12}^{}] \right ) {\rm Re}[\hat{\kappa}_{12}^{} +
\hat{\kappa}_{13}^{}] \sin2 \Delta_{31}^{} \sin \left ( V_{CC}^{} L
\right )
\nonumber\\
& & + 4 \left [ \left ( \frac{2 \Delta_{21}^{}}{3 V_{CC}^{} L}
\right )^2 - \left | \hat{\kappa}_{12}^{} \right |^2
\right ] \sin^2 \left ( \frac{V_{CC}^{} L}{2} \right ) \nonumber\\
& & \underline{+ 2 \left ( \frac{2 \Delta_{21}^{}}{3 V_{CC}^{} L}
{\rm Im}[\hat{\kappa}_{12}^{} + \hat{\kappa}_{13}^{}] - {\rm
Im}[\hat{\kappa}_{12}^{} \hat{\kappa}_{13}^{*}] \right ) \sin 2
\Delta_{31}^{} \sin^2 \left ( \frac{V_{CC}^{} L}{2} \right )} \nonumber\\
& & \underline{+ 2 \left ( \frac{2 \Delta_{21}^{}}{3 V_{CC}^{} L}
{\rm Im}[\hat{\kappa}_{12}^{} + \hat{\kappa}_{13}^{}] - {\rm
Im}[\hat{\kappa}_{12}^{} \hat{\kappa}_{13}^{*}] \right ) + \sin^2
\Delta_{31}^{} \sin \left (
V_{CC}^{} L \right )} \nonumber\\
& & \underline{+ 2 {\rm Im}[\hat{\kappa}_{12}^{}
\hat{\kappa}_{13}^{*}] \sin 2 \Delta_{31}^{} + \frac{8
\Delta_{21}^{}}{3 V_{CC}^{} L}  {\rm Im}[\hat{\kappa}_{12}^{}] \sin
\left ( V_{CC}^{} L \right )} \; ,\\
\nonumber\\
\tilde{P}_{e \tau}^{} & \; = \; & 4 |\hat{\kappa}_{13}^{}|^2 - \left
( 3 |\hat{\kappa}_{13}^{}|^2 - |\hat{\kappa}_{12}^{}|^2 + 2 {\rm
Re}[\hat{\kappa}_{12}^{} \hat{\kappa}_{13}^{*}] \right )
\sin^2 \Delta_{31}^{} \nonumber\\
& & - 4 \left ( \frac{2 \Delta_{21}^{}}{3 V_{CC}^{} L} - {\rm
Re}[\hat{\kappa}_{13}^{}] \right ) {\rm Re}[\hat{\kappa}_{12}^{} +
\hat{\kappa}_{13}^{}] \sin^2 \Delta_{31}^{} \sin^2 \left (
\frac{V_{CC}^{} L}{2} \right )
\nonumber\\
& & - \left ( \frac{2 \Delta_{21}^{}}{3 V_{CC}^{} L} - {\rm
Re}[\hat{\kappa}_{13}^{}] \right ) {\rm Re}[\hat{\kappa}_{12}^{} +
\hat{\kappa}_{13}^{}] \sin2 \Delta_{31}^{} \sin \left ( V_{CC}^{} L
\right )
\nonumber\\
& & + 4 \left [ \left ( \frac{2 \Delta_{21}^{}}{3 V_{CC}^{} L}
\right )^2 - \left | \hat{\kappa}_{13}^{} \right |^2 \right ] \sin^2
\left ( \frac{V_{CC}^{} L}{2} \right ) \nonumber\\
& & \underline{- 2 \left ( \frac{2 \Delta_{21}^{}}{3 V_{CC}^{} L}
{\rm Im}[\hat{\kappa}_{12}^{} + \hat{\kappa}_{13}^{}] - {\rm
Im}[\hat{\kappa}_{12}^{} \hat{\kappa}_{13}^{*}] \right ) \sin 2
\Delta_{31}^{} \sin^2 \left ( \frac{V_{CC}^{} L}{2} \right )} \nonumber\\
& & \underline{- 2 \left ( \frac{2 \Delta_{21}^{}}{3 V_{CC}^{} L}
{\rm Im}[\hat{\kappa}_{12}^{} + \hat{\kappa}_{13}^{}] - {\rm
Im}[\hat{\kappa}_{12}^{} \hat{\kappa}_{13}^{*}] \right ) \sin^2
\Delta_{31}^{} \sin \left (
V_{CC}^{} L \right )} \nonumber\\
& & \underline{- 2 {\rm Im}[\hat{\kappa}_{12}^{}
\hat{\kappa}_{13}^{*}] \sin 2 \Delta_{31}^{} - \frac{8
\Delta_{21}^{}}{3 V_{CC}^{} L}  {\rm Im}[\hat{\kappa}_{13}^{}] \sin
\left (
V_{CC}^{} L \right )} \; , \\
\nonumber\\
\tilde{P}_{\mu \tau}^{} & \; = \; & 4 |\hat{\kappa}_{23}^{}|^2 +
\left [ 1 + 4 \left ( \frac{1}{3} \Delta_{21}^{} + 2 V_{NC}^{} L
{\rm Re}[\hat{\kappa}_{23}^{}] \right )^2 - |\hat{\kappa}_{12}^{} +
\hat{\kappa}_{13}^{}|^2 \right. \nonumber\\
& & \left. - 4 |\hat{\kappa}_{23}^{}|^2 - 4\left ( {\rm
Im}[\hat{\kappa}_{23}^{}] \right )^2 - \frac{\left ( V_{CC}^{} L
\right )^2}{\Delta_{31}^{}} \left ( \epsilon_{b}^{} -\epsilon_{c}^{}
\right )^2 \right ] \sin^2 \Delta_{31}^{} \nonumber\\
& & - \left [ 2 \left ( \frac{1}{3} \Delta_{21}^{} + 2 V_{NC}^{} L
{\rm Re}[\hat{\kappa}_{23}^{}] \right ) + \frac{8 \Delta_{21}^{2}}{9
V_{CC}^{} L} - 2 V_{NC}^{} L \left ( \epsilon_{b}^{} +
\epsilon_{c}^{} \right ) {\rm Re}[\hat{\kappa}_{23}^{}] \right.
\nonumber\\
& & \left. + 2 \left ( V_{CC}^{} - V_{NC}^{} \right ) L {\rm
Re}[\hat{\kappa}_{12}^{} \hat{\kappa}_{13}^{*}] + \frac{\left (
V_{CC}^{} L \right )^2}{\Delta_{31}^{}} \left [ \left (
\epsilon_{b}^{} - \epsilon_{c}^{} \right )^2 + 4 \left ( {\rm Im}
[\hat{\kappa}_{23}^{}] \right )^2 \right ] \right ]
\sin2\Delta_{31}^{}
\nonumber\\
& & - 2 \left ( \frac{4 \Delta_{21}^{2}}{9 \left ( V_{CC}^{} L
\right )^2} + \frac{2 \Delta_{21}^{}}{3 V_{CC}^{} L} {\rm
Re}[\hat{\kappa}_{12}^{} - \hat{\kappa}_{13}^{}] - {\rm
Re}[\hat{\kappa}_{12}^{} \hat{\kappa}_{13}^{*}] \right ) \nonumber\\
& & \cdot \left [ 2 \sin^2 \Delta_{31}^{} \sin^2 \left (
\frac{V_{CC}^{} L}{2} \right ) + \frac{1}{2} \sin2 \Delta_{31}^{}
\sin \left ( V_{CC}^{} L \right ) \right ] \nonumber\\
& & + 4 \left ( \frac{1}{3} \Delta_{21}^{} + 2 V_{CC}^{} L {\rm
Re}[\hat{\kappa}_{23}^{}] \right )^2 \underline{+ \left [ \left ( 2
- \epsilon_{b}^{} - \epsilon_{c}^{} \begin{matrix} \cr \end{matrix}
- \frac{4 \left ( V_{NC}^{} L \right )^2}{\Delta_{31}^{}} {\rm
Re}[\hat{\kappa}_{23}^{}] \right. \right.} \nonumber\\
& & \underline{\left. \left. - \frac{2 \sqrt{2} \Delta_{21}^{}
V_{NC}^{} L}{3 \Delta_{31}^{} \left (
V_{CC}^{} - 2 \Delta_{31}^{} \right )} \right ) {\rm Im}[\hat{\kappa}_{23}^{}] - {\rm Im}[\hat{\kappa}_{12}^{} \hat{\kappa}_{13}^{*}] \right ] \sin 2 \Delta_{31}^{}}\nonumber\\
& & \underline{+ 2 \left ( \frac{2 \Delta_{21}^{}}{3 V_{CC}^{} L}
{\rm Im}[\hat{\kappa}_{12}^{} + \hat{\kappa}_{13}^{}] - {\rm
Im}[\hat{\kappa}_{12}^{} \hat{\kappa}_{13}^{*}] \right ) \sin 2
\Delta_{31}^{} \sin^2 \left ( \frac{V_{CC}^{} L}{2} \right )}
\nonumber\\
& & \underline{+ 2 \left ( \frac{2 \Delta_{21}^{}}{3 V_{CC}^{} L}
{\rm Im}[\hat{\kappa}_{12}^{} + \hat{\kappa}_{13}^{}] - {\rm
Im}[\hat{\kappa}_{12}^{} \hat{\kappa}_{13}^{*}] \right ) \sin^2
\Delta_{31}^{} \sin \left ( V_{CC}^{} L \right )}
\nonumber\\
& & \underline{ + 16 {\rm Im}[\hat{\kappa}_{23}^{}] \left (
\frac{1}{3} \Delta_{21}^{} + 2 V_{NC}^{} L {\rm
Re}[\hat{\kappa}_{23}^{}] \right ) \sin^2 \Delta_{31}^{}} \nonumber\\
& & \underline{ - 8 {\rm Im}[\hat{\kappa}_{23}^{}] \left (
\frac{1}{3} \Delta_{21}^{} + 2 V_{NC}^{} L {\rm
Re}[\hat{\kappa}_{23}^{}] \right )} \; .
\end{eqnarray}
In order to obtain the probabilities of anti-neutrino oscillations
$\bar{\nu}_{\alpha}^{} \rightarrow \bar{\nu}_{\beta}^{}$, one
needs to simultaneously change the signs of $V_{CC}^{}$,
$V_{NC}^{}$ and the terms underlined in the expressions of
$\tilde{P}_{\alpha \beta}^{}$.

Different from the case in vacuum, the terms of ${\cal
O}(\hat{\kappa}_{ij}^{})$ which can be strongly enhanced by large
$L$ appear not only in the expression of $\tilde{P}_{\mu \tau}^{}$
but also in those of $\tilde{P}_{\mu \mu}^{}$ and $\tilde{P}_{\tau
\tau}^{}$. If we omit the terms of ${\cal
O}(\hat{\kappa}_{ij}^{2})$, then we get very concise formulas for
$\tilde{P}_{\mu \mu}^{}$, $\tilde{P}_{\tau \tau}^{}$ and
$\tilde{P}_{\mu \tau}^{}$:
\begin{eqnarray}
\tilde{P}_{\mu \mu}^{} & \; = \; & 1 - \sin^2 \Delta_{31}^{} +
2\left ( \frac{1}{3} \Delta_{21}^{} + 2 V_{NC}^{} L {\rm
Re}[\hat{\kappa}_{23}^{}] \right ) \sin2\Delta_{31}^{} \; ,
\\
\tilde{P}_{\tau \tau}^{} & \; = \; & 1 - \sin^2 \Delta_{31}^{} + 2
\left ( \frac{1}{3} \Delta_{21}^{} + 2 V_{NC}^{} L {\rm
Re}[\hat{\kappa}_{23}^{}] \right ) \sin2\Delta_{31}^{} \; ,
\\
\tilde{P}_{\mu \tau}^{} & \; = \; & \sin^2 \Delta_{31}^{} + 2 \left
( \frac{1}{3} \Delta_{21}^{} + 2 V_{NC}^{} L {\rm
Re}[\hat{\kappa}_{23}^{}] \right ) \sin2\Delta_{31}^{} + 2 {\rm
Im}[\hat{\kappa}_{23}^{}] \sin2\Delta_{31}^{} \; .
\end{eqnarray}

We carry out a numerical analysis to show the difference between the
corrections from the non-unitary parameter $\hat{\kappa}_{23}^{}$
and the corrections from the nonzero $\theta_{13}^{}$ to the
neutrino oscillation probabilities. We compare between two special
cases: Case I, we consider a unitary and nearly tri-bimaximal mixing
matrix with nonzero $\theta_{13}^{}$, as $V_{0}^{}$ given in Eq. (2)
with $\theta_{12}^{} = \arcsin{\displaystyle \frac{1}{\sqrt{3}}}$
and $\theta_{23}^{} = 45^{\circ}$; Case II, we consider the
non-unitary mixing matrix as shown in Eq. (6). The inputs of our
numerical calculations are $\Delta m_{21}^{2} = 8.0 \times 10^{-5} ~
{\rm eV}^2$, $\Delta m_{31}^{2} = \pm ~ 2.5 \times 10^{-3} ~ {\rm
eV}^2$ (sign ``$+$" for the normal hierarchy and ``$-$" for the
inverted hierarchy), the matter density $\rho = 2.7 ~ {\rm g} / {\rm
cm}^3$. And in Case I we choose $\theta_{13}^{} = 10^{\circ}$, in
Case II we choose $\epsilon_{a}^{} = - 5.5 \times 10^{-3}$,
$\epsilon_{b}^{} = - 5.0 \times 10^{-3}$, $\epsilon_{c}^{} = - 5.0
\times 10^{-3}$, $\hat{\kappa}_{12}^{} = 3.5 \times 10^{-5} \cdot
e^{i\frac{\pi}{4}}$, $\hat{\kappa}_{13}^{} = 8.0 \times 10^{-3}
\cdot e^{i\frac{\pi}{4}}$ and $|\hat{\kappa}_{23}^{}| = 5.0 \times
10^{-3}$. Our numerical analysis is independent of our analytical
results given above, but it confirms the results of our analytical
approximations. As for the analytical approximations for Case I, one
may refer to \cite{0402175}, in which there exists a complete set of
series expansion for three-flavor neutrino oscillation probabilities
in matter in terms of small $\theta_{13}^{}$ and $\alpha \equiv
\Delta m_{21}^{2} / \Delta m_{31}^{2}$.

Fig. 1 shows the effective mass differences in matter as functions
of the neutrino beam energy $E_{\nu}^{}$ in Case I and Case II for
both the normal and the inverted hierarchies. We can clearly see
that the mass squared differences are strongly magnified if
$E_{\nu}^{}$ is large (or equivalently if the matter density is
large). An interesting point is that in Case II the effective mass
difference $\Delta \tilde{m}_{32}^{2}$ can reach zero at around
$E_{\nu}^{} = 12 ~ {\rm GeV}$ in the normal hierarchy case while in
Case I the nonzero $\sin \theta_{13}^{}$ ensures the nonzero value
of $\Delta \tilde{m}_{32}^{2}$. If we choose a nonzero
$\theta_{13}^{}$ for Case II, we will get similar curves as those of
Case I. We find that although the non-unitary parameter
$\displaystyle \frac{1}{\sqrt{2}} \left ( \hat{\kappa}_{12}^{} +
\hat{\kappa}_{13}^{} \right )$ plays a very similar role as $\sin
\theta_{13}^{} e_{}^{- i \delta}$ in the expressions of neutrino
oscillation probabilities in vacuum, it has very different effects
from $\sin \theta_{13}^{} e_{}^{- i \delta}$ in matter. This finding
provides us with a possibility of distinguishing the nonzero
$\theta_{13}^{}$ from the non-unitary parameters in $V$.

Taking account of Eqs. (27) to (29) in Ref. \cite{0402175}, we can
easily see that the Dirac phase $\delta$ dose not appear in the
expressions of the eigenvalues of $\tilde{\cal H}$. However, in the
non-unitary case, all the effective mass squared differences contain
the terms proportional to $E_{\nu}^{} V_{NC}^{} {\rm
Re}[\hat{\kappa}_{23}^{}]$ which is relevant to ${\rm
arg}(\hat{\kappa}_{23}^{})$ for fixed $|\hat{\kappa}_{23}^{}|$. Fig.
2 shows the effective mass differences in matter as functions of the
Dirac phase $\delta$ in Case I or the phase of
$\hat{\kappa}_{23}^{}$ in Case II for both mass hierarchies, where
we have chosen $E_{\nu}^{} = 50 ~ {\rm GeV}$. We find that the
correction from the term $E_{\nu}^{} V_{NC}^{} {\rm
Re}[\hat{\kappa}_{23}^{}]$ can be around $10^{-5} ~ {\rm eV}^2$ in
this situation.

Fig. 3 tells us how the probabilities of neutrino oscillations
$\nu_{\mu}^{} \rightarrow \nu_{\mu}^{}$ and $\nu_{\mu}^{}
\rightarrow \nu_{\tau}^{}$ in matter are modified with the
changing of the baseline $L$, where we have chosen $\delta =
\displaystyle \frac{\pi}{4}$ and ${\rm arg} (\hat{\kappa}_{23}^{})
= \displaystyle \frac{\pi}{4}$. We can clearly see from the figure
that the probability $\tilde{P}_{\mu \tau}^{}$ ($\tilde{P}_{\mu
\mu}^{}$) can be largely enhanced (depressed) by a long baseline
if the matter effect is taken into account. At the baseline $L =
4000 ~ {\rm km}$, $\tilde{P}_{\mu \tau}^{}$ is about $10^{-3}$.
Fig. 4 shows the probability $\tilde{P}_{\mu \tau}^{}$ and
$\tilde{P}_{\mu \mu}^{}$ as functions of the Dirac phase $\delta$
in Case I or the phase of $\hat{\kappa}_{23}^{}$ in Case II for
both mass hierarchies with the baselines $L = 1000 ~ {\rm km}$ and
$L = 4000 ~ {\rm km}$. We find that if $\tilde{P}_{\mu \tau}^{}$
can be measured to the level of $10^{-4}$ at $4000 ~ {\rm km}$
from the source and $|\hat{\kappa}_{23}^{}|$ can be well measured,
${\rm arg}(\hat{\kappa}_{23}^{})$ may convincingly be determined.
Another point worthwhile to point out is that from Eqs. (33) to
(35) we find that these three probabilities have approximate
sign[$\Delta m_{31}^{2}$] - ${\rm arg}(\hat{\kappa}_{23}^{})$
degeneracy, which means $\tilde{P}_{\mu \mu, ~ \tau \tau, ~ \mu
\tau}^{}(\Delta m_{31}^{2}, {\rm arg}(\hat{\kappa}_{23}^{}))
\approx \tilde{P}_{\mu \mu, ~ \tau \tau, ~ \mu \tau}^{}(- \Delta
m_{31}^{2}, - {\rm arg}(\hat{\kappa}_{23}^{}))$. This degeneracy
is broken by the term $\displaystyle \frac{2}{3} \Delta_{21}^{}
\sin 2\Delta_{31}^{}$, which increases with the increase of the
baseline $L$. Fig. 4 shows that this degeneracy breaking can reach
$10^{-3}$ at $L = 4000 ~ {\rm km}$ if the energy of neutrinos is
taken to be $E_{\nu}^{} = 50 ~ {\rm GeV}$.

We admit that it is very difficult to measure the transition
probabilities to the accuracy of $10^{-3}$ or even $10^{-4}$ in
the present or forthcoming neutrino oscillation experiments. Given
the small effects of unitarity violation, however, our numerical
results can at least serve to illustrate how sensitive an
ambitious long-baseline neutrino experiment should be to this kind
of new physics. It is worth remarking two positive aspects of
searching for the non-unitarity of $V$ in the $\nu_{\mu}^{}$
disappearance or $\nu_{\tau}^{}$ appearance experiments. First,
the signatures of the non-unitarity can be strongly enhanced by
the matter effects, and thus a high energy and a very long
baseline are essential to detect appreciable effects of the
non-unitarity in the neutrino oscillation experiments. As for
neutrinos of energy around $50 ~ {\rm GeV}$, a baseline much
longer than $4000 ~ {\rm km}$ may have much better sensitivity to
the non-unitary parameters in the neutrino mixing matrix, in which
case the varying of the terrestrial matter density need to be
taken into account \cite{varying}. Second, the dependency of the
non-unitary parameters $\hat{\kappa}_{ij}^{}$ on the energy of the
neutrino beam is different from other neutrino mixing parameters,
and thus measuring the energy dependency of the transition
probabilities will help to identify the small effects of the
non-unitarity.

\section{Constructing ``deformed unitarity triangles"}

In the unitary limit, the neutrino mixing matrix $V$ which relates
the neutrino mass eigenstates $(\nu_{1}^{}, \nu_{2}^{},
\nu_{3}^{})$ to the neutrino flavor eigenstates $(\nu_{e}^{},
\nu_{\mu}^{}, \nu_{\tau}^{})$ is unitary. The unitarity implies
\begin{eqnarray}
\Delta_{\tau}^{} : & ~~~ & V_{e 1}^{} V_{\mu 1}^{*} + V_{e 2}^{}
V_{\mu 2}^{*} + V_{e 3}^{} V_{\mu 3}^{*} \; = \; 0 \; ,
\nonumber \\
\Delta_{\mu}^{} : & ~~~ & V_{e 1}^{} V_{\tau 1}^{*} + V_{e 2}^{}
V_{\tau 2}^{*} + V_{e 3}^{} V_{\tau 3}^{*} \; = \; 0 \; ,
\nonumber \\
\Delta_{e}^{} : & ~~~ & V_{\mu 1}^{} V_{\tau 1}^{*} + V_{\mu 2}^{}
V_{\tau 2}^{*} + V_{\mu 3}^{} V_{\tau 3}^{*} \; = \; 0 \; ,
\nonumber \\
\Delta_{3}^{} : & ~~~ & V_{e 1}^{} V_{e 2}^{*} + V_{\mu 1}^{} V_{\mu
2}^{*} + V_{\tau 1}^{} V_{\tau 2}^{*} \; = \; 0 \; ,
\nonumber \\
\Delta_{2}^{} : & ~~~ & V_{e 1}^{} V_{e 3}^{*} + V_{\mu 1}^{} V_{\mu
3}^{*} + V_{\tau 1}^{} V_{\tau 3}^{*} \; = \; 0 \; ,
\nonumber \\
\Delta_{1}^{} : & ~~~ & V_{e 2}^{} V_{e 3}^{*} + V_{\mu 2}^{} V_{\mu
3}^{*} + V_{\tau 2}^{} V_{\tau 3}^{*} \; = \; 0 \; .
\end{eqnarray}
In the complex plane, Eq. (36) corresponds to six unitarity
triangles \cite{unitarity triangle} denoted as $\Delta_{\tau}^{}$,
$\Delta_{\mu}^{}$, $\Delta_{e}^{}$, $\Delta_{3}^{}$, $\Delta_{2}^{}$
and $\Delta_{1}^{}$ respectively. The area of each unitarity
triangle is $\frac{1}{2} |{\cal J}|$, where ${\cal J}$ is the
Jarlskog invariant measure of CP violation for the unitary MNS
mixing matrix. If there is no CP violation (e.g., the tri-bimaximal
mixing), the unitarity triangles shrink to segments. In other words,
introducing the ``unitarity triangles" provides a geometric way to
describe CP violation (which can be determined by the area of each
triangle) by measuring the CP-conserving quantities (the sides of
the triangles).

If the mixing matrix $V$ is non-unitary, the orthogonal relations in
Eq. (36) are in general not satisfied and the unitarity triangles in
the complex plane are in general open. Note that every two vector
sides can determine one triangle in the complex plane, of which
twice the area corresponds to one of the nine Jarlskog invariants.
To be explicit, the Jarlskog invariant $J_{\alpha \beta}^{ij}$
equals to twice the area of the triangle determined by $V_{\alpha
i}^{} V_{\beta i}^{*}$ and $V_{\alpha j}^{} V_{\beta j}^{*}$ (or
$V_{\alpha i}^{} V_{\alpha j}^{*}$ and $V_{\beta i}^{} V_{\beta
j}^{*}$). Apparently, if all the six triangles are closed, these
nine Jarlskog invariants are all equal (to ${\cal J}$).

As for our special scenario, the CP-violating effects are attributed
to the phases of $\hat{\kappa}_{12}^{}$, $\hat{\kappa}_{13}^{}$ and
$\hat{\kappa}_{23}^{}$. Thus we are able to determine those
unitarity-violating parameters by constructing those ``deformed
unitarity triangles". In Appendix B, we give all the eighteen sides
of the six deformed triangles. Table I shows the ratios of two sides
of $\Delta_{\tau}^{}$, $\Delta_{\mu}^{}$ and $\Delta_{e}^{}$ to the
first order of $\hat{\kappa}_{13}^{}$ and $\hat{\kappa}_{23}^{}$.
Here we omit the smallest parameter $\hat{\kappa}_{12}^{}$.

\begin{table}[h]
\caption{Ratios of the absolute values of the sides of
$\Delta_{\tau}^{}$, $\Delta_{\mu}^{}$ and $\Delta_{e}^{}$, to the
first order in $\hat{\kappa}_{13}^{}$ and $\hat{\kappa}_{23}^{}$.
The smallest parameter $\hat{\kappa}_{12}^{}$ is omitted.}
\begin{center}
\begin{tabular}{c|l}
\hline\\[-12pt]
~ $\Delta_{\tau}^{}$ ~ & ~ $\displaystyle
\frac{|S_{2}^{}|}{|S_{1}^{}|} = \left |
\frac{V_{e2}^{}V_{\mu2}^{*}}{V_{e1}^{}V_{\mu1}^{*}} \right | \approx
|1 -
\frac{3}{2} \hat{\kappa}_{13}^{}|$ ~ \\[12pt]
& ~ $\displaystyle \frac{|S_{3}^{}|}{|S_{1}^{}|} = \left |
\frac{V_{e3}^{}V_{\mu3}^{*}}{V_{e1}^{}V_{\mu1}^{*}} \right | \approx
\frac{|S_{3}^{}|}{|S_{2}^{}|} = \left |
\frac{V_{e3}^{}V_{\mu3}^{*}}{V_{e1}^{}V_{\mu1}^{*}} \right | \approx
\frac{3}{2} |\hat{\kappa}_{13}^{}|$ ~ \\[12pt]
\hline\\[-12pt]
~ $\Delta_{\mu}^{}$ ~ & ~ $\displaystyle
\frac{|S_{2}^{}|}{|S_{1}^{}|} = \left |
\frac{V_{e2}^{}V_{\tau2}^{*}}{V_{e1}^{}V_{\tau1}^{*}} \right |
\approx |1 -
\frac{9}{2} \hat{\kappa}_{13}^{}|$ ~ \\[12pt]
& ~ $\displaystyle \frac{|S_{3}^{}|}{|S_{1}^{}|} = \left |
\frac{V_{e3}^{}V_{\tau3}^{*}}{V_{e1}^{}V_{\tau1}^{*}} \right |
\approx \frac{|S_{3}^{}|}{|S_{2}^{}|} = \left |
\frac{V_{e3}^{}V_{\tau3}^{*}}{V_{e1}^{}V_{\tau1}^{*}} \right |
\approx \frac{3}{2} |\hat{\kappa}_{13}^{}|$ ~ \\[12pt]
\hline\\[-12pt]
~ $\Delta_{e}^{}$ ~ & ~ $\displaystyle \frac{|S_{1}^{}|}{|S_{3}^{}|}
= \left | \frac{V_{\mu2}^{}V_{\mu2}^{*}}{V_{\mu1}^{}V_{\tau1}^{*}}
\right | \approx \frac{1}{3} |1 -
4 \hat{\kappa}_{23}^{} + \hat{\kappa}_{13}^{}|$ ~ \\[12pt]
& ~ $\displaystyle \frac{|S_{2}^{}|}{|S_{3}^{}|} = \left |
\frac{V_{\mu3}^{}V_{\mu3}^{*}}{V_{\mu1}^{}V_{\tau1}^{*}} \right |
\approx \frac{1}{3} |1 -
4 \hat{\kappa}_{23}^{} - \hat{\kappa}_{13}^{}|$ ~ \\[12pt]
& ~ $\displaystyle \frac{|S_{2}^{}|}{|S_{1}^{}|} = \left |
\frac{V_{\mu3}^{}V_{\mu3}^{*}}{V_{\mu1}^{}V_{\tau1}^{*}} \right |
\approx 2 |1 -
3 \hat{\kappa}_{13}^{}|$ ~ \\[12pt]
\hline
\end{tabular}
\end{center}
\end{table}

One can find from Appendix B that for $\Delta_{\tau}^{}$ we have
$S_{1}^{} + S_{2}^{} + S_{3}^{} \approx 2 \hat{\kappa}_{12}^{}
\approx 0$, therefore $\Delta_{\tau}^{}$ is almost closed. From
Table I we find that in our special scenario, if the ratio
$|S_{3}^{}| / |S_{1}^{}|$ of $\Delta_{\tau}^{}$ can be well
measured, $|\hat{\kappa}_{13}^{}|$ can be determined. Combined with
the ratio $|S_{2}^{}| / |S_{1}^{}|$, the phase of
$\hat{\kappa}_{13}^{}$ can also be obtained. The results can be and
should be checked by the measurements of $|S_{3}^{}| / |S_{1}^{}|$
and $|S_{2}^{}| / |S_{1}^{}|$ of $\Delta_{\mu}^{}$, which is a
validation of our scenario. In addition, constructing the triangle
$\Delta_{e}^{}$ can give us the correlation between
$|\hat{\kappa}_{23}^{}|$ and ${\rm arg}(\hat{\kappa}_{23}^{})$ if
$\hat{\kappa}_{13}^{}$ is fixed.

\section{Further discussions}

In this section we would like to present a short discussion about
the well-accepted seesaw mechanism \cite{seesaw}, of which the
unitary violation of the mixing matrix $V$ is a general consequence.
We show here how to incorporate our parametrization into a generic
type-II seesaw model. In the type-II seesaw model \cite{type II
seesaw} which contains $n$ right-handed neutrinos, the neutrino mass
terms can be written as
\begin{eqnarray}
-{\cal L}_{\rm mass} \; = \; \frac{1}{2} ~ \overline{\left(
\begin{matrix} \nu^{}_{\rm L} & N^c_{\rm R} \end{matrix} \right)} ~ \left(
\begin{matrix} M^{}_{\rm L} & M^{}_{\rm D} \cr M^T_{\rm D} &
M^{}_{\rm R}
\end{matrix} \right) \left( \begin{matrix} \nu^c_{\rm L} \cr N^{}_{\rm R}
\end{matrix} \right) ~ + ~ {\rm h.c.} \; ,
\end{eqnarray}
where $\nu^c_{\rm L} \equiv C \overline{\nu^{}_{\rm L}}^T$ with $C$
being the charge conjugation matrix, likewise for $N^c_{\rm R}$.
Here, $M^{}_{\rm L}$ is a $3 \times 3$ matrix, $M^{}_{\rm D}$ is a
$3 \times n$ matrix and $M^{}_{\rm R}$ is a $n \times n$ matrix. The
overall $(n+3) \times (n+3)$ neutrino mass matrix in ${\cal
L}^{}_{\rm mass}$, denoted as ${\cal M}$, can be diagonalized by the
unitary transformation ${\cal U}^\dagger {\cal M} {\cal U}^* =
\widehat{\cal M}$; or explicitly,
\begin{eqnarray}
\left( \begin{matrix} V & R \cr S & U \end{matrix} \right)^\dagger
\left(
\begin{matrix} M^{}_{\rm L} & M^{}_{\rm D} \cr M^T_{\rm D} &
M^{}_{\rm R} \end{matrix} \right) \left( \begin{matrix} V & R \cr S
& U \end{matrix} \right)^* = \left(
\begin{matrix} \widehat{M}^{}_\nu & {\bf 0} \cr {\bf 0} &
\widehat{M}^{}_{\rm N} \end{matrix} \right) \; ,
\end{eqnarray}
where $\widehat{M}^{}_\nu = {\rm diag}\left ( m^{}_1, m^{}_2, m^{}_3
\right )$ and $\widehat{M}^{}_{\rm N} = {\rm diag}\left ( M^{}_1,
\cdots, M^{}_n \right )$ with $m^{}_i$ and $M^{}_i$ being the light
and heavy Majorana neutrino masses, respectively. Note that the
submatrices $V$, $U$, $R$ and $S$ are all non-unitary. Suppose the
mass eigenstates and the flavor eigenstates of the charged leptons
are identical. In the basis of mass states, the standard
charged-current interactions of neutrinos can be written as
\begin{eqnarray}
-{\cal L}^{}_{\rm cc} \; = \; \frac{g}{\sqrt{2}} \left[ \overline{
\left( \begin{matrix} e & \mu & \tau \end{matrix} \right)^{}_{\rm
L}} ~V \gamma^\mu \left(
\begin{matrix} \nu^{}_1 \cr \nu^{}_2 \cr \nu^{}_3 \end{matrix} \right)^{}_{\rm
L} W^-_{\mu} + \overline{\left( \begin{matrix} e & \mu & \tau
\end{matrix} \right)^{}_{\rm L}} ~R \gamma^\mu \left( \begin{matrix}
N^{}_1 \cr \vdots \cr N^{}_n \end{matrix} \right)^{}_{\rm L} W^-_\mu
\right] ~ + ~ {\rm h.c.} \; .
\end{eqnarray}
One can find that the matrix $R$ describes the strength of the
charged-current interaction between the charged leptons and the
heavy neutrinos. In addition, the deviation of $V$ from unitary is
also characterized by $R$, since the unitarity of ${\cal U}$
requires $V V^\dagger + R R^\dagger = {\bf 1}$ \cite{R}.

In fact the diagonalization of ${\cal M}$ can be divided into two
steps by decomposing ${\cal U}$ into the product of two unitary
matrices ${\cal W}$ and $\cal V$:
\begin{eqnarray}
{\cal V}^\dagger {\cal W}^\dagger \left(  \begin{matrix} M^{}_{\rm
L} & M^{}_{\rm D} \cr M^T_{\rm D} & M^{}_{\rm R}
\end{matrix} \right) {\cal W}^* {\cal V}^* \equiv {\cal V}^\dagger
\left(
\begin{matrix}  M^{}_\nu & {\bf 0} \cr {\bf 0} & M^{}_{\rm
N} \end{matrix} \right) {\cal V}^* = \left(
 \begin{matrix}  \widehat{M}^{}_\nu & {\bf 0} \cr {\bf 0} &
\widehat{M}^{}_{\rm N} \end{matrix} \right) \; ,
\end{eqnarray}
where ${\cal W}$ and ${\cal V}$ take the general forms
\begin{eqnarray}
{\cal W} = \left(  \begin{matrix}  U^{}_L & B \cr C & U^{}_R
\end{matrix} \right) \; , ~~~ {\cal V} = \left(  \begin{matrix}
V^{}_L & {\bf 0} \cr {\bf 0} & V^{}_R \end{matrix} \right) \; .
\end{eqnarray}
The matrices $U^{}_L$, $B$, $C$ and $U^{}_R$ are in general
non-unitary, but $V^{}_L$ and $V^{}_R$ are unitary matrices. We can
easily find that: $V = U^{}_L V^{}_L$, $R = B V^{}_R$. Let's count
the degrees of freedom of these matrices. The $(n + 3) \times (n +
3)$ unitary matrix contains $(n + 3)^2$ degrees of freedom. Suppose
freedom left for ${\cal W}$ is $(n+3)^2 - 3^2 - n^2 = 2 \times 3$
the parameters in $V^{}_L$ and $V^{}_R$ are all free, the degrees of
$\times n$. An ansatz made for ${\cal W}$ in Ref. \cite{Grimus} is
to suppose that $B$ is an arbitrary $3 \times n$ matrix which
contains just $2 \times 3n$ degrees of freedom and then parametrize
${\cal W}$ as
\begin{eqnarray}
{\cal W} = \left(  \begin{matrix}  \sqrt{{\bf 1} -B B^\dagger} & B
\cr -B^\dagger & \sqrt{{\bf 1} -B^\dagger B}
\end{matrix} \right) \; .
\end{eqnarray}

Comparing Eqs. (38), (40), (41) and (42) with the parametrization
$V = H \cdot V_{0}^{}$ shown in Section I, we can simply choose:
$V_{0}^{} = V_{L}^{}$, then we can find that $H = U_{L}^{} =
\sqrt{{\bf 1} -B B^\dagger} = \sqrt{{\bf 1} -B V^{}_R
V^{\dagger}_R B^\dagger_{}} = \sqrt{{\bf 1} -R R^\dagger}$ which
contains the same degrees of freedom as the Hermitian matrix in
Eq. (1). That is to say in the parametrization we have chosen, the
unitarity-violating parameters in $H$ can be expressed as the
functions of $RR_{}^{\dagger}$, where $R$ is of the order of
$M_{D}^{} / M_{R}^{}$. In other words, the deviation from the
unitarity which is described by $\mathbf{\epsilon}$ in Eq. (4) is
of the order of $\left ( \frac{M_{D}^{}}{M_{R}^{}} \right )^2$.

In the canonical seesaw scenarios, the light neutrino masses are
attributed to the leading-order contribution $M_{\nu}^{} = -
M_{D}^{} M_{R}^{-1} M_{D}^{T}$ (type-I seesaw) or $M_{\nu}^{} =
M_{L}^{} - M_{D}^{} M_{R}^{-1} M_{D}^{T}$ (type-II seesaw). In
order to obtain the light neutrino mass scale $m_{\nu}^{} \sim 0.1
~ {\rm eV}$, the mass scale of right-handed Majorana neutrinos is
expected to be as high as $m_{R}^{} \sim 10^{14} ~ {\rm GeV}$. In
such a case, one finds $M_{D}^{} / M_{R}^{} \sim 10^{-12}$ or
equivalently $\mathbf{\epsilon} \sim 10^{-24}$, which is too small
to be detected. The possible ways out have recently been discussed
(see, e.g., Refs. \cite{Kersten} and \cite{Chao}). For instance,
one may first impose certain flavor symmetries on the textures of
$M_{D}^{}$, $M_{R}^{}$ and $M_{L}^{}$ to guarantee $M_{D}^{}
M_{R}^{-1} M_{D}^{T} = 0$ (type-I seesaw) or $M_{L}^{} - M_{D}^{}
M_{R}^{-1} M_{D}^{T} = 0$ (type-II seesaw) and then introduce
slight perturbations to them so as to produce the tiny light
neutrino masses. As a consequence, the mass scale of three light
neutrinos in this approach is essentially decoupled from the
magnitude of $R$. In the models proposed in Refs. \cite{Kersten,
Chao}, for example, the right-handed Majorana neutrinos are
assumed to be around the TeV scale such that $M_{D}^{} / M_{R}^{}
\sim {\cal{O}} ~ (10^{-1})$ holds and the elements of
$\mathbf{\epsilon}$ can be as large as $\sim {\cal{O}} ~
(10^{-2})$, just close to their upper bounds constrained by
current experimental data. These kinds of seesaw models will be
tested at the LHC or ILC \cite{TeV}; and one of their low-energy
consequences, which is just the unitarity violation of $V$ under
discussion, will also be tested in the long-baseline neutrino
oscillation experiments.

\section{Summary}

As we are about to enter an era of high precision neutrino physics,
a general and comprehensive analysis of the non-unitary neutrino
mixing matrix makes sense and will be useful for phenomenological
explanations of future measurements and tests of type-I and type-II
seesaw models. In this paper we have investigated a new pattern of
the neutrino mixing matrix which is the product of an arbitrary
Hermitian matrix and the well-known tri-bimaximal mixing matrix.
Starting with this non-unitary mixing matrix, we have presented a
complete set of series expansion formulas for neutrino oscillation
probabilities both in vacuum and in matter of constant density. We
have carried out a numerical analysis to emphasize the importance of
matter effects in the measurements of the non-unitary parameters and
in distinguishing their effects from the effects induced by small
$\theta_{13}^{}$. We find that measuring the probability of
$\nu_{\mu}^{} \rightarrow \nu_{\tau}^{}$ or $\nu_{\mu}^{}
\rightarrow \nu_{\mu}^{}$ oscillation with large neutrino energy
(e.g., $\sim 50 ~ {\rm GeV}$) and a relatively long baseline (e.g.,
several thousand km) is a viable way to detect those non-unitary
parameters. We have also discussed the possibility of determining
the small non-unitary perturbations and the extra CP-violating
phases by constructing the ``deformed unitarity triangles". Finally
we have shown that our parametrization of the non-unitary mixing
matrix can be naturally incorporated into the generic type-II seesaw
model.

\begin{acknowledgements}

I would like to thank Prof. Z.Z. Xing for sharing valuable ideas
with me and enlightening me on this subject. I am greatly indebted
to him for polishing up the manuscript with many suggestions and
corrections. I am also grateful to W. Chao for patient and useful
discussions. This work was supported in part by the National Natural
Science Foundation of China.

\end{acknowledgements}

\begin{appendix}

\section{Diagonalizing the Hamiltonian in matter by using the perturbation theory}
In this appendix we use the perturbation theory to diagonalize the
effect Hamiltonian $\tilde{{\cal H}} = U_{}^{} \tilde{E}
U_{}^{\dagger}$ in matter of constant density. In the series
expansion of $\tilde{\cal{H}}$, we regard $\hat{\kappa}_{12}^{}$,
$\hat{\kappa}_{13}^{}$, $\hat{\kappa}_{23}^{}$, $\epsilon_{a}^{}$,
$\epsilon_{b}^{}$, $\epsilon_{c}^{}$ and $\alpha \equiv \Delta
m_{21}^{2} / \Delta m_{31}^{2}$ as the small parameters of the same
order, and perform the diagonalization to the second order of them.
Now we are going to diagonalize the effective Hamiltonian
$\tilde{{\cal H}} = E + V_{}^{T} A V_{}^{*}$.

First we draw $E_{1}^{} \cdot \mathbf{1}$ (which contributes only a
pure phase $e_{}^{-i E_{1}^{} L}$ to the oscillation amplitude and
nothing to the oscillation probabilities) out of $\tilde{{\cal H}}$
and rewrite $\tilde{{\cal H}}$ as
\begin{equation}
\tilde{\cal{H}} \; = \; E' + V_{}^{T} A V_{}^{*} \; = \;
\frac{1}{2E_{\nu}^{}} \cdot {\rm diag} \left ( 0, \alpha, \Delta
m_{31}^{2} \right ) + V_{}^{T} A V_{}^{*} \; ,
\end{equation}
where $A \equiv {\rm diag} \left ( V_{CC}^{} - V_{NC}^{}, -
V_{NC}^{}, - V_{NC}^{} \right )$. We find that it is easier to
perform the diagonalization of $\tilde{{\cal H'}} \equiv V_{0}^{}
\tilde{{\cal H}} V_{0}^{\dagger} = V_{0}^{} E' V_{0}^{\dagger} +
H_{}^{T} A H_{}^{*}$, with
\begin{equation}
V_{0}^{} E V_{0}^{\dagger} \; = \; \frac{\Delta
m_{31}^{2}}{2E_{\nu}^{}} \left [ \frac{1}{2} \left ( \begin{matrix}
~ 0 ~ & ~ 0 ~ & ~ 0 ~ \cr ~ 0 ~ & ~ 1 ~ & ~ 1 ~ \cr ~ 0 ~ & ~ 1 ~ &
~ 1 ~
\end{matrix} \right ) + \displaystyle \frac{\alpha}{3} \left ( \begin{matrix} ~ 1 &
~ 1 & - 1 \cr ~ 1 & ~ 1 & - 1 \cr - 1 & - 1 & ~ 1
\end{matrix} \right ) \right ] \; ,
\end{equation}
and
\begin{eqnarray}
H_{}^{T} A H_{}^{*} & \; = \; & \left ( \mathbf{1} +
\mathbf{\epsilon} \right )_{}^{T} A \left ( \mathbf{1} +
\mathbf{\epsilon} \right )_{}^{*} \nonumber\\ \nonumber\\
& \; = \; & V_{CC}^{} \left [\left ( \begin{matrix} ~ 1 ~ & & \cr &
0 & \cr & & ~ 0 ~
\end{matrix} \right ) + \mathbf{\epsilon}_{}^{*}
\left ( \begin{matrix} ~ 1 ~ & & \cr & 0 & \cr & & ~ 0 ~
\end{matrix} \right ) + \left ( \begin{matrix} ~ 1 ~ & & \cr &
0 & \cr & & ~ 0 ~
\end{matrix} \right ) \mathbf{\epsilon}_{}^{*} + \mathbf{\epsilon}_{}^{*}
\left ( \begin{matrix} ~ 1 ~ & & \cr & 0 & \cr & & ~ 0 ~
\end{matrix} \right ) \mathbf{\epsilon}_{}^{*} \right ] \nonumber\\ \nonumber\\
& & - V_{NC}^{} \left ( 2 \mathbf{\epsilon}_{}^{*} +
{\mathbf{\epsilon}_{}^{*}}^2\right ) - V_{NC}^{} \cdot \mathbf{1} \;
,
\end{eqnarray}
The term $- V_{NC}^{} \cdot \mathbf{1}$ can be neglected for the
same reason as omitting the term $E_{1}^{} \cdot \mathbf{1}$. Then
we will diagonalize $\tilde{{\cal H'}}$ with $\tilde{{\cal H'}} =
W_{}^{} \tilde{E'} W_{}^{\dagger}$ using the perturbation theory. We
can easily find that $\tilde{E} = \tilde{E'} + E_{1}^{} - V_{CC}^{}$
and $U = V_{0}^{\dagger} W$. The matrix $X$, which describes the
mixing in matter, can be expressed as $X = V U_{}^{*} = H V_{0}^{}
V_{0}^{\dagger} W = H W$, which is also non-unitary.

We write
\begin{equation}
\tilde{\cal{H'}} \; = \; \tilde{\cal{H'}}_{}^{(0)} +
\tilde{\cal{H'}}_{}^{(1)} + \tilde{\cal{H'}}_{}^{(2)} \; ,
\end{equation}
where
\begin{eqnarray}
\tilde{\cal{H'}}_{}^{(0)} & \; = \; & \frac{\Delta
m_{31}^{2}}{4E_{\nu}^{}} \left ( \begin{matrix} ~ 0 ~ & ~ 0 ~ & ~ 0
~ \cr ~ 0 ~ & ~ 1 ~ & ~ 1 ~ \cr ~ 0 ~ & ~ 1 ~ & ~ 1 ~
\end{matrix} \right ) + V_{CC}^{} \left ( \begin{matrix} ~ 1 ~ & & \cr &
~ 0 ~ & \cr & & ~ 0 ~
\end{matrix} \right ) \; , \\
\nonumber\\
\tilde{\cal{H'}}_{}^{(1)} & \; = \; & \frac{1}{3} \cdot \frac{\Delta
m_{21}^{2}}{2E_{\nu}^{}} \left (
\begin{matrix} ~ 1 & ~ 1 & - 1 \cr ~ 1 & ~ 1 & - 1 \cr - 1 & - 1 & ~
1
\end{matrix} \right ) + V_{CC}^{} \left (
\begin{matrix} 2 \epsilon_{a}^{} & \hat{\kappa}_{12}^{*} &
\hat{\kappa}_{13}^{*} \cr \hat{\kappa}_{12}^{} & 0 & 0 \cr
\hat{\kappa}_{13}^{} & 0 & 0
\end{matrix} \right ) - 2 V_{NC}^{} \left (
\begin{matrix} \epsilon_{a}^{} & \hat{\kappa}_{12}^{*} &
\hat{\kappa}_{13}^{*} \cr \hat{\kappa}_{12}^{} & \epsilon_{b}^{} &
\hat{\kappa}_{23}^{*} \cr \hat{\kappa}_{13}^{} &
\hat{\kappa}_{23}^{} & \epsilon_{c}^{}
\end{matrix} \right ) \; , \\
\nonumber\\
\tilde{\cal{H'}}_{}^{(2)} & \; = \; & V_{CC}^{} \left (
\begin{matrix} \epsilon_{a}^{2} & \epsilon_{a}^{} \hat{\kappa}_{12}^{*} &
\epsilon_{a}^{} \hat{\kappa}_{13}^{*} \cr \epsilon_{a}^{}
\hat{\kappa}_{12}^{} & |\hat{\kappa}_{12}^{}|^2 &
\hat{\kappa}_{12}^{} \hat{\kappa}_{13}^{*} \cr \epsilon_{a}^{}
\hat{\kappa}_{13}^{} & \hat{\kappa}_{12}^{*} \hat{\kappa}_{13}^{} &
|\hat{\kappa}_{13}^{}|^2
\end{matrix} \right ) \nonumber\\ \nonumber\\
& & - V_{NC}^{} \left (
\begin{matrix} \epsilon_{a}^{2} + |\hat{\kappa}_{12}^{}|^2 + |\hat{\kappa}_{13}^{}|^2 & \left ( \epsilon_{a}^{} + \epsilon_{b}^{} \right )
\hat{\kappa}_{12}^{*} + \hat{\kappa}_{13}^{*} \hat{\kappa}_{23}^{} &
\left ( \epsilon_{a}^{} + \epsilon_{c}^{} \right )
\hat{\kappa}_{13}^{*} + \hat{\kappa}_{12}^{*} \hat{\kappa}_{23}^{}
\cr \left ( \epsilon_{a}^{} + \epsilon_{b}^{} \right )
\hat{\kappa}_{12}^{} + \hat{\kappa}_{13}^{} \hat{\kappa}_{23}^{*} &
\epsilon_{b}^{2} + |\hat{\kappa}_{12}^{}|^2 +
|\hat{\kappa}_{23}^{}|^2 & \left ( \epsilon_{b}^{} + \epsilon_{c}^{}
\right ) \hat{\kappa}_{23}^{*} + \hat{\kappa}_{12}^{}
\hat{\kappa}_{13}^{*} \cr \left ( \epsilon_{a}^{} + \epsilon_{c}^{}
\right ) \hat{\kappa}_{13}^{} + \hat{\kappa}_{12}^{}
\hat{\kappa}_{23}^{} & \left ( \epsilon_{b}^{} + \epsilon_{c}^{}
\right ) \hat{\kappa}_{23}^{} + \hat{\kappa}_{12}^{*}
\hat{\kappa}_{13}^{} & \epsilon_{c}^{2} + |\hat{\kappa}_{13}^{}|^2 +
|\hat{\kappa}_{23}^{}|^2
\end{matrix} \right ) \; . ~~~~~~
\end{eqnarray}
For the eigenvalues and the eigenvectors, we also write
$\tilde{E'}_{i}^{} \; = \; \tilde{E'}_{i}^{(0)} +
\tilde{E'}_{i}^{(1)} + \tilde{E'}_{i}^{(2)}$ and $v_{i}^{} \; = \;
v_{i}^{(0)} + v_{i}^{(1)} + v_{i}^{(2)}$ (for $i = 1,2,3$). The
unitary matrix $W = \left ( v_{1}^{}, v_{2}^{}, v_{3}^{} \right )$.

$\tilde{\cal{H'}}_{}^{(0)}$ can be easily diagonalized. The
eigenvalues and the eigenvectors of $\tilde{\cal{H'}}_{}^{(0)}$ are:
\begin{equation}
\tilde{E'}_{1}^{(0)} \; = \; V_{CC}^{} \; , ~~~ \tilde{E'}_{2}^{(0)}
\; = \; 0 \; , ~~~ \tilde{E'}_{i}^{(0)} \; = \; \frac{\Delta
m_{31}^{2}}{2E_{\nu}^{}} \; ;
\end{equation}
and
\begin{equation}
v_{1}^{(0)} \; = \; \left ( 1, 0, 0 \right )^T \; , ~~~ v_{2}^{(0)}
\; = \; \left ( 0, \frac{1}{\sqrt{2}}, - \frac{1}{\sqrt{2}} \right
)^T \; , ~~~ v_{3}^{(0)} \; = \; \left ( 0, \frac{1}{\sqrt{2}},
\frac{1}{\sqrt{2}} \right )^T \; .
\end{equation}
Then, the first and the second order corrections to the eigenvalues
are given by
\begin{eqnarray}
\tilde{E'}_{i}^{(1)} & \; = \; & \tilde{\cal{H'}}_{ii}^{(1)} \; , \\
\tilde{E'}_{i}^{(2)} & \; = \; & \tilde{\cal{H'}}_{ii}^{(2)} +
\sum_{j \neq i}^{} \frac{ \left | \tilde{\cal{H'}}_{ji}^{(1)} \right
|^2}{\tilde{E'}_{i}^{(0)} - \tilde{E'}_{j}^{(0)}} \; ;
\end{eqnarray}
and the corrections to the eigenvectors are calculated by
\begin{eqnarray}
v_{i}^{(1)} & \; = \; & \sum_{j \neq i}^{} \frac{
\tilde{\cal{H'}}_{ji}^{(1)}
}{\tilde{E'}_{i}^{(0)} - \tilde{E'}_{j}^{(0)}} \cdot v_{j}^{(0)} \; , \\
\nonumber\\
v_{i}^{(2)} & \; = \; &  \sum_{j \neq i}^{} \left [ \frac{
\tilde{\cal{H'}}_{ji}^{(2)}}{\tilde{E'}_{i}^{(0)} -
\tilde{E'}_{j}^{(0)}} + \sum_{k \neq i}^{}
\frac{\tilde{\cal{H'}}_{jk}^{(1)} \tilde{\cal{H'}}_{ki}^{(1)}}{
\left ( \tilde{E'}_{i}^{(0)} - \tilde{E'}_{j}^{(0)} \right ) \left (
\tilde{E'}_{i}^{(0)} - \tilde{E'}_{k}^{(0)} \right ) } -
\frac{\tilde{\cal{H'}}_{ii}^{(1)} \tilde{\cal{H'}}_{jk}^{(1)}}{
\left ( \tilde{E'}_{i}^{(0)} - \tilde{E'}_{j}^{(0)} \right )^2 }
\right ] \cdot v_{j}^{(0)} \nonumber\\ \nonumber\\
& & - \frac{1}{2} \left [ \sum_{j \neq i}^{} \frac{ \left |
\tilde{\cal{H'}}_{ji}^{(1)} \right |^2}{ \left (
\tilde{E'}_{i}^{(0)} - \tilde{E'}_{j}^{(0)} \right )^2} \right ]
\cdot v_{i}^{(0)} \; ,
\end{eqnarray}
where $\tilde{\cal{H'}}_{ij}^{(n)} \equiv {v_{i}^{(0)}}^{\dagger}
\tilde{\cal{H'}}_{}^{(n)} v_{j}^{(0)}$.

Inserting Eqs. (A8), (A9) into Eqs. (A10) and (A12), we obtain
\begin{eqnarray}
\tilde{E'}_{1}^{(0)} & \; = \; & \frac{1}{3} \cdot \frac{\Delta
m_{21}^{2}}{2E_{\nu}^{}} + 2 \epsilon_{a}^{} \left ( V_{CC}^{} - V_{NC}^{} \right ) \; , \\
\tilde{E'}_{2}^{(0)} & \; = \; & \frac{2}{3} \cdot \frac{\Delta
m_{21}^{2}}{2E_{\nu}^{}} - V_{NC}^{} \left ( \epsilon_{b}^{} + \epsilon_{c}^{} + 2 {\rm Re} [\hat{\kappa}_{23}^{}] \right ) \; , \\
\tilde{E'}_{3}^{(0)} & \; = \; & - V_{NC}^{} \left ( \epsilon_{b}^{}
+ \epsilon_{c}^{} - 2 {\rm Re} [\hat{\kappa}_{23}^{}] \right ) \; ;
\end{eqnarray}
and
\begin{equation}
W_{}^{(1)} \; = \; \left ( \begin{matrix} 0 & - \displaystyle
\frac{2 \sqrt{2} \Delta_{21}^{}}{3 V_{CC}^{} L} & 0 \cr ~
\displaystyle \frac{2 \Delta_{21}^{}}{3 V_{CC}^{} L} &\displaystyle
\frac{V_{NC}^{}}{2 \sqrt{2} \Delta_{31}^{}} \left ( \epsilon_{b}^{}
- \epsilon_{c}^{} + 2 {\rm Im}[\hat{\kappa}_{23}^{}] \right ) & -
\displaystyle \frac{V_{NC}^{}}{2 \sqrt{2} \Delta_{31}^{}} \left (
\epsilon_{b}^{} - \epsilon_{c}^{} - 2 {\rm Im}[\hat{\kappa}_{23}^{}]
\right ) \cr - \displaystyle \frac{2 \Delta_{21}^{}}{3 V_{CC}^{} L}
& \displaystyle \frac{V_{NC}^{}}{2 \sqrt{2} \Delta_{31}^{}} \left (
\epsilon_{b}^{} - \epsilon_{c}^{} + 2 {\rm Im}[\hat{\kappa}_{23}^{}]
\right ) & ~ \displaystyle \frac{V_{NC}^{}}{2 \sqrt{2}
\Delta_{31}^{}} \left ( \epsilon_{b}^{} - \epsilon_{c}^{} - 2 {\rm
Im}[\hat{\kappa}_{23}^{}] \right )
\end{matrix} \right ) \; ,
\end{equation}
where $\Delta_{ij}^{} \equiv \Delta m_{ij}^{2}L/(4E_{\nu}^{})$
with $\Delta m_{ij}^{2} \equiv m_{i}^{2} - m_{j}^{2}$ (for $ij =
21, 31, 32$), and the terms proportional to $\left ( V_{CC}^{} - 2
V_{NC}^{} \right )$ are omitted.

We can further calculate $X = H W$ to the first order in
$\hat{\kappa}_{12}^{}$, $\hat{\kappa}_{13}^{}$,
$\hat{\kappa}_{23}^{}$, $\epsilon_{a}^{}$, $\epsilon_{b}^{}$,
$\epsilon_{c}^{}$ and $\alpha$:
\begin{equation}
X \; \approx \; \left ( \begin{matrix} 1 + \epsilon_{a}^{} & -
\displaystyle \frac{2 \sqrt{2} \Delta_{21}^{}}{3 V_{CC}^{} L} +
\frac{\hat{\kappa}_{12}^{} - \hat{\kappa}_{13}^{}}{\sqrt{2}} &
\displaystyle \frac{\hat{\kappa}_{12}^{} +
\hat{\kappa}_{13}^{}}{\sqrt{2}} \cr ~ \displaystyle \frac{2
\Delta_{21}^{}}{3 V_{CC}^{} L} + \hat{\kappa}_{12}^{*} &
\displaystyle ~ \frac{b - \hat{\kappa}_{23}^{}}{\sqrt{2}} +
\frac{V_{NC}^{} \left ( \epsilon_{b}^{} - \epsilon_{c}^{} + 2 {\rm
Im}[\hat{\kappa}_{23}^{}] \right )}{2 \sqrt{2} \Delta_{31}^{}} &
\displaystyle \frac{ b + \hat{\kappa}_{23}^{}}{\sqrt{2}} -
\frac{V_{NC}^{} \left ( \epsilon_{b}^{} - \epsilon_{c}^{} - 2 {\rm
Im}[\hat{\kappa}_{23}^{}] \right )}{2 \sqrt{2} \Delta_{31}^{}} \cr -
\displaystyle \frac{2 \Delta_{21}^{}}{3 V_{CC}^{} L} +
\hat{\kappa}_{13}^{*} & \displaystyle - \frac{c -
\hat{\kappa}_{23}^{}}{\sqrt{2}} + \frac{V_{NC}^{} \left (
\epsilon_{b}^{} - \epsilon_{c}^{} + 2 {\rm Im}[\hat{\kappa}_{23}^{}]
\right )}{2 \sqrt{2} \Delta_{31}^{}} & \displaystyle \frac{c +
\hat{\kappa}_{23}^{}}{\sqrt{2}} + \frac{V_{NC}^{} \left (
\epsilon_{b}^{} - \epsilon_{c}^{} - 2 {\rm Im}[\hat{\kappa}_{23}^{}]
\right )}{2 \sqrt{2} \Delta_{31}^{}}
\end{matrix} \right ) \; .
\end{equation}

In this paper we do not order the eigenvalues of $\tilde{\cal H}$
according to their magnitude and the mass spectrum. This ordering
does not change the oscillation probabilities. In this appendix we
give the results of $\tilde{E'}_{i}^{(0)}$,
$\tilde{E'}_{i}^{(1)}$, $W_{}^{(0)}$ and $W_{}^{(1)}$, which are
enough for calculating the probabilities $P_{ee}^{}$, $P_{e
\mu}^{}$ and $P_{e \tau}^{}$ to the second order.
$\tilde{E'}_{i}^{(2)}$ and $v_{i}^{(2)}$, which are not shown
here, only correct the terms of the second order in $P_{\mu
\mu}^{}$, $P_{\mu \tau}^{}$ and $P_{\tau \tau}^{}$.

\section{Sides of six ``deformed unitarity triangles"}
Sides of six ``deformed unitarity triangles" to the first order in
$\hat{\kappa}_{12}^{}$, $\hat{\kappa}_{13}^{}$,
$\hat{\kappa}_{23}^{}$, $\epsilon_{a}^{}$, $\epsilon_{b}^{}$ and
$\epsilon_{c}^{}$.
\begin{itemize}
\item $\Delta_{\tau}^{}$:
\begin{eqnarray}
S_{1}^{} & \; = \; & V_{e 1}^{} V_{\mu 1}^{*} \; \approx \; -
\frac{1}{3} - \frac{1}{3} \left ( \epsilon_{a}^{} + \epsilon_{b}^{}
\right ) + \frac{1}{6} \left ( 5 \hat{\kappa}_{12}^{} -
\hat{\kappa}_{13}^{} + 2 \hat{\kappa}_{23}^{*} \right )
\; ,\\
S_{2}^{} & \; = \; & V_{e 2}^{} V_{\mu 2}^{*} \; \approx \;
\frac{1}{3} + \frac{1}{3} \left ( \epsilon_{a}^{} + \epsilon_{b}^{}
\right ) + \frac{1}{3} \left ( 2 \hat{\kappa}_{12}^{} -
\hat{\kappa}_{13}^{} - \hat{\kappa}_{23}^{*} \right )
\; ,\\
S_{3}^{} & \; = \; & V_{e 3}^{} V_{\mu 3}^{*} \; \approx \;
\frac{1}{2} \left ( \hat{\kappa}_{12}^{} + \hat{\kappa}_{13}^{}
\right ) \; ,
\end{eqnarray}
with $S_{1}^{} + S_{2}^{} + S_{3}^{} \approx 2
\hat{\kappa}_{12}^{}$.
\item $\Delta_{\mu}^{}$:
\begin{eqnarray}
S_{1}^{} & \; = \; & V_{e 1}^{} V_{\tau 1}^{*} \; \approx \;
\frac{1}{3} + \frac{1}{3} \left ( \epsilon_{a}^{} + \epsilon_{c}^{}
\right ) + \frac{1}{6} \left ( 5 \hat{\kappa}_{13}^{} -
\hat{\kappa}_{12}^{} - 2 \hat{\kappa}_{23}^{*} \right )
\; ,\\
S_{2}^{} & \; = \; & V_{e 2}^{} V_{\tau 2}^{*} \; \approx \; -
\frac{1}{3} - \frac{1}{3} \left ( \epsilon_{a}^{} + \epsilon_{c}^{}
\right ) + \frac{1}{3} \left ( 2 \hat{\kappa}_{13}^{} -
\hat{\kappa}_{12}^{} + \hat{\kappa}_{23}^{*} \right )
\; ,\\
S_{3}^{} & \; = \; & V_{e 3}^{} V_{\tau 3}^{*} \; \approx \;
\frac{1}{2} \left ( \hat{\kappa}_{12}^{} + \hat{\kappa}_{13}^{}
\right ) \; ,
\end{eqnarray}
with $S_{1}^{} + S_{2}^{} + S_{3}^{} \approx 2
\hat{\kappa}_{13}^{}$.
\item $\Delta_{e}^{}$:
\begin{eqnarray}
S_{1}^{} & \; = \; & V_{\mu 1}^{} V_{\tau 1}^{*} \; \approx \; -
\frac{1}{6} - \frac{1}{6} \left ( \epsilon_{b}^{} + \epsilon_{c}^{}
\right ) + \frac{1}{3} \left ( \hat{\kappa}_{23}^{} +
\hat{\kappa}_{12}^{*} - \hat{\kappa}_{13}^{} \right )
\; ,\\
S_{2}^{} & \; = \; & V_{\mu 2}^{} V_{\tau 2}^{*} \; \approx \; -
\frac{1}{3} - \frac{1}{3} \left ( \epsilon_{b}^{} + \epsilon_{c}^{}
\right ) + \frac{1}{3} \left ( 2 \hat{\kappa}_{23}^{} -
\hat{\kappa}_{12}^{*} + \hat{\kappa}_{13}^{} \right )
\; ,\\
S_{3}^{} & \; = \; & V_{\mu 3}^{} V_{\tau 3}^{*} \; \approx \;
\frac{1}{2} + \frac{1}{2} \left ( \epsilon_{b}^{} + \epsilon_{c}^{}
\right ) +  \hat{\kappa}_{23}^{} \; ,
\end{eqnarray}
with $S_{1}^{} + S_{2}^{} + S_{3}^{} \approx 2
\hat{\kappa}_{23}^{}$.
\item $\Delta_{3}^{}$:
\begin{eqnarray}
S_{1}^{} & \; = \; & V_{e 1}^{} V_{e 2}^{*} \; \approx \;
\frac{\sqrt{2}}{3} \left ( 1 + 2 \epsilon_{a}^{} \right ) -
\frac{1}{3 \sqrt{2}} \left ( {\rm Re}[\hat{\kappa}_{12}^{} -
\hat{\kappa}_{13}^{}] -3 i {\rm Im}[\hat{\kappa}_{12}^{} -
\hat{\kappa}_{13}^{}] \right )
\; ,\\
S_{2}^{} & \; = \; & V_{\mu 1}^{} V_{\mu 2}^{*} \; \approx \; -
\frac{1}{3 \sqrt{2}} \left ( 1 + 2 \epsilon_{b}^{} \right ) +
\frac{1}{3 \sqrt{2}} \left ( {\rm Re}[\hat{\kappa}_{12}^{} + 2
\hat{\kappa}_{23}^{}] -3 i {\rm Im}[\hat{\kappa}_{12}^{}] \right )
\; ,\\
S_{3}^{} & \; = \; & V_{\tau 1}^{} V_{\tau 2}^{*} \; \approx \; -
\frac{1}{3 \sqrt{2}} \left ( 1 + 2 \epsilon_{c}^{} \right ) -
\frac{1}{3 \sqrt{2}} \left ( {\rm Re}[\hat{\kappa}_{13}^{} - 2
\hat{\kappa}_{23}^{}] -3 i {\rm Im}[\hat{\kappa}_{13}^{}] \right )
\; ,
\end{eqnarray}
with $S_{1}^{} + S_{2}^{} + S_{3}^{} \approx - \displaystyle
\frac{\sqrt{2}}{3} \left ( 2 \epsilon_{a}^{} - \epsilon_{b}^{} -
\epsilon_{c}^{} \right ) + \displaystyle \frac{\sqrt{2}}{3} \left (
{\rm Re}[2 \hat{\kappa}_{23}^{} + \hat{\kappa}_{12}^{} -
\hat{\kappa}_{13}^{}] -3 i {\rm Im}[\hat{\kappa}_{12}^{} -
\hat{\kappa}_{13}^{}] \right )$.
\item $\Delta_{2}^{}$:
\begin{eqnarray}
S_{1}^{} & \; = \; & V_{e 1}^{} V_{e 3}^{*} \; \approx \;
\frac{1}{\sqrt{3}} \left ( \hat{\kappa}_{12}^{*} +
\hat{\kappa}_{13}^{*} \right )
\; ,\\
S_{2}^{} & \; = \; & V_{\mu 1}^{} V_{\mu 3}^{*} \; \approx \; -
\frac{1}{2 \sqrt{3}} \left ( 1 + 2 \epsilon_{b}^{} \right ) +
\frac{1}{\sqrt{3}} \left ( \hat{\kappa}_{12}^{*} + i {\rm
Im}[\hat{\kappa}_{23}^{}] \right )
\; ,\\
S_{3}^{} & \; = \; & V_{\tau 1}^{} V_{\tau 3}^{*} \; \approx \;
\frac{1}{2 \sqrt{3}} \left ( 1 + 2 \epsilon_{c}^{} \right ) +
\frac{1}{\sqrt{3}} \left ( \hat{\kappa}_{13}^{*} + i {\rm
Im}[\hat{\kappa}_{23}^{}] \right ) \; ,
\end{eqnarray}
with $S_{1}^{} + S_{2}^{} + S_{3}^{} \approx \displaystyle
\frac{1}{2 \sqrt{3}} \left ( \epsilon_{b}^{} - \epsilon_{c}^{}
\right ) - \displaystyle \frac{2}{\sqrt{3}} \left (
\hat{\kappa}_{12}^{*} + \hat{\kappa}_{13}^{*} + i {\rm
Im}[\hat{\kappa}_{23}^{}] \right )$.
\item $\Delta_{1}^{}$:
\begin{eqnarray}
S_{1}^{} & \; = \; & V_{e 2}^{} V_{e 3}^{*} \; \approx \;
\frac{1}{\sqrt{6}} \left ( \hat{\kappa}_{12}^{*} +
\hat{\kappa}_{13}^{*} \right )
\; ,\\
S_{2}^{} & \; = \; & V_{\mu 2}^{} V_{\mu 3}^{*} \; \approx \;
\frac{1}{\sqrt{6}} \left ( 1 + 2 \epsilon_{b}^{} \right ) +
\frac{1}{\sqrt{6}} \left ( \hat{\kappa}_{12}^{*} - 2 i {\rm
Im}[\hat{\kappa}_{23}^{}] \right )
\; ,\\
S_{3}^{} & \; = \; & V_{\tau 2}^{} V_{\tau 3}^{*} \; \approx \; -
\frac{1}{\sqrt{6}} \left ( 1 + 2 \epsilon_{c}^{} \right ) +
\frac{1}{\sqrt{6}} \left ( \hat{\kappa}_{13}^{*} - 2 i {\rm
Im}[\hat{\kappa}_{23}^{}] \right ) \; ,
\end{eqnarray}
with $S_{1}^{} + S_{2}^{} + S_{3}^{} \approx - \displaystyle
\sqrt{\frac{2}{3}} \left ( \epsilon_{b}^{} - \epsilon_{c}^{} \right
) + \displaystyle \sqrt{\frac{2}{3}} \left ( \hat{\kappa}_{12}^{*} +
\hat{\kappa}_{13}^{*} - 2 i {\rm Im}[\hat{\kappa}_{23}^{}] \right
)$.
\end{itemize}

\end{appendix}

\newpage

\begin{figure}[h]
\begin{center}
\vspace{9cm}
\includegraphics[bbllx=6.5cm, bblly=6.0cm, bburx=15.0cm, bbury=14.2cm,%
width=7.9cm, height=7.9cm, angle=0, clip=0]{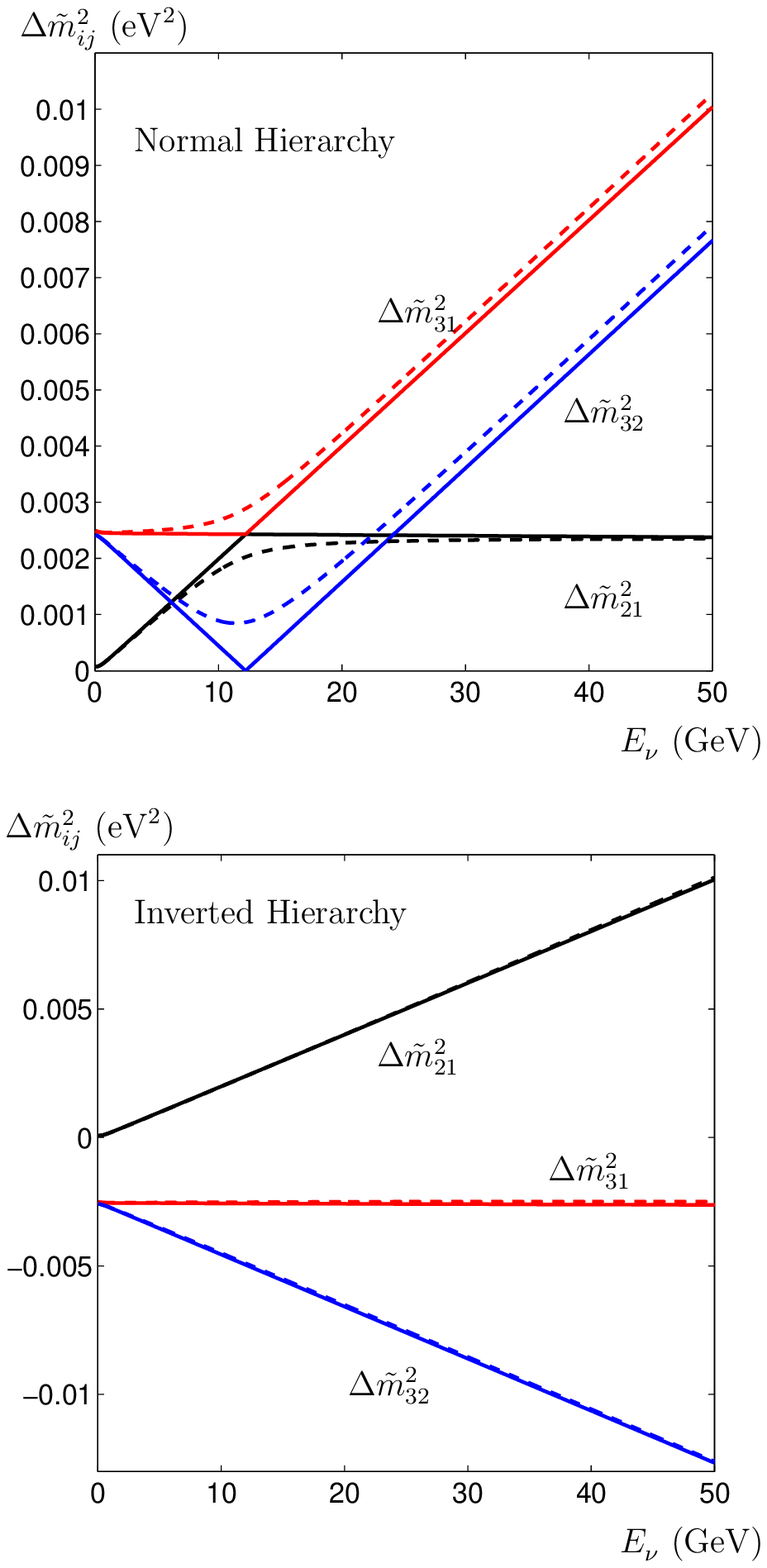}
\vspace{2.4cm}\caption{The effective mass-squared differences in
matter as functions of the neutrino beam energy $E_{\nu}^{}$ in Case
I (the unitary case, represented by dashed lines) and Case II (the
non-unitary case, represented by solid lines) for both the normal
(the first plot) and the inverted (the second plot) hierarchies.}
\end{center}
\end{figure}

\begin{figure}
\begin{center}
\vspace{11.5cm} \subfigure[~Normal hierarchy]{
\includegraphics[bbllx=5.2cm, bblly=4.8cm, bburx=15.0cm, bbury=14.2cm,%
width=7.6cm, height=7.6cm, angle=0, clip=0]{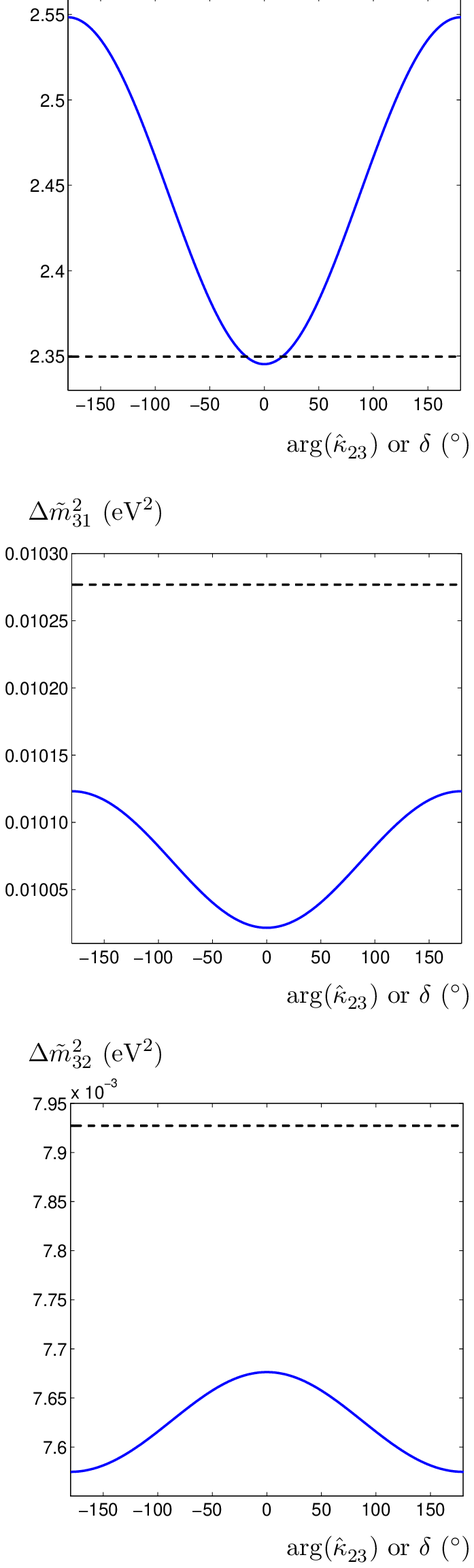}}
\subfigure[~Inverted hierarchy]{
\includegraphics[bbllx=5.2cm, bblly=4.8cm, bburx=15.0cm, bbury=14.2cm,%
width=7.6cm, height=7.6cm, angle=0, clip=0]{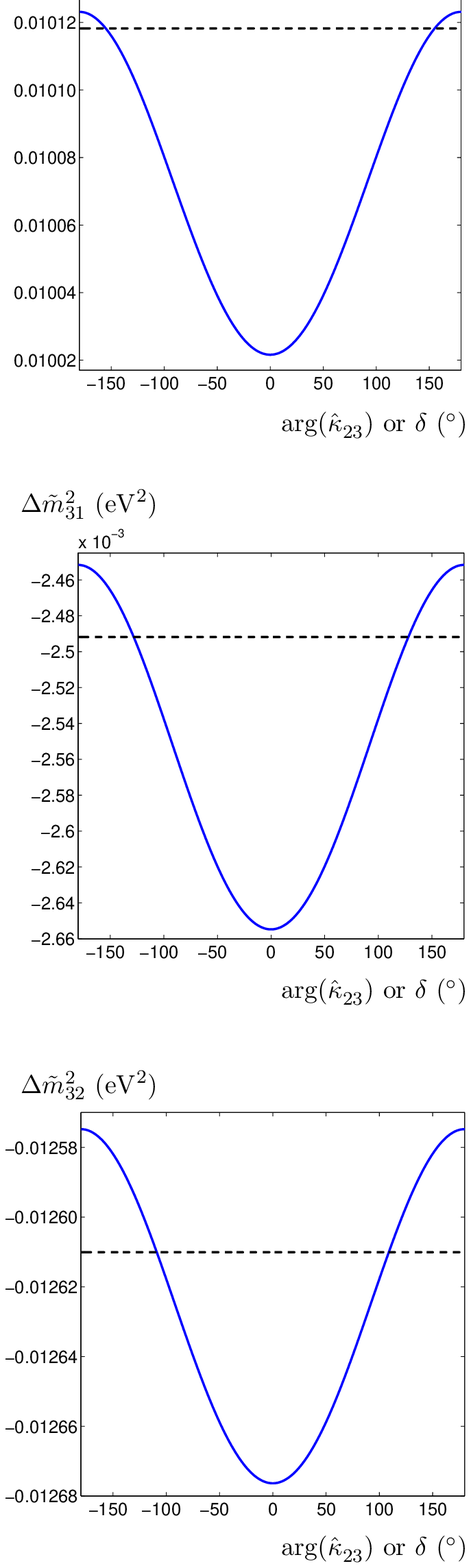}}
\vspace{0.3cm}\caption{The effective mass-squared differences in
matter as functions of the Dirac phase $\delta$ in Case I (the
unitary case, represented by dashed lines) or the phase of
$\hat{\kappa}_{23}^{}$ in Case II (the non-unitary case, represented
by solid lines) for both mass hierarchies, where we choose
$E_{\nu}^{} = 50 ~ {\rm GeV}$.}
\end{center}
\end{figure}

\begin{figure}[h]
\begin{center}
\vspace{9cm}
\includegraphics[bbllx=6.5cm, bblly=6.0cm, bburx=15.0cm, bbury=14.2cm,%
width=7.0cm, height=7.0cm, angle=0, clip=0]{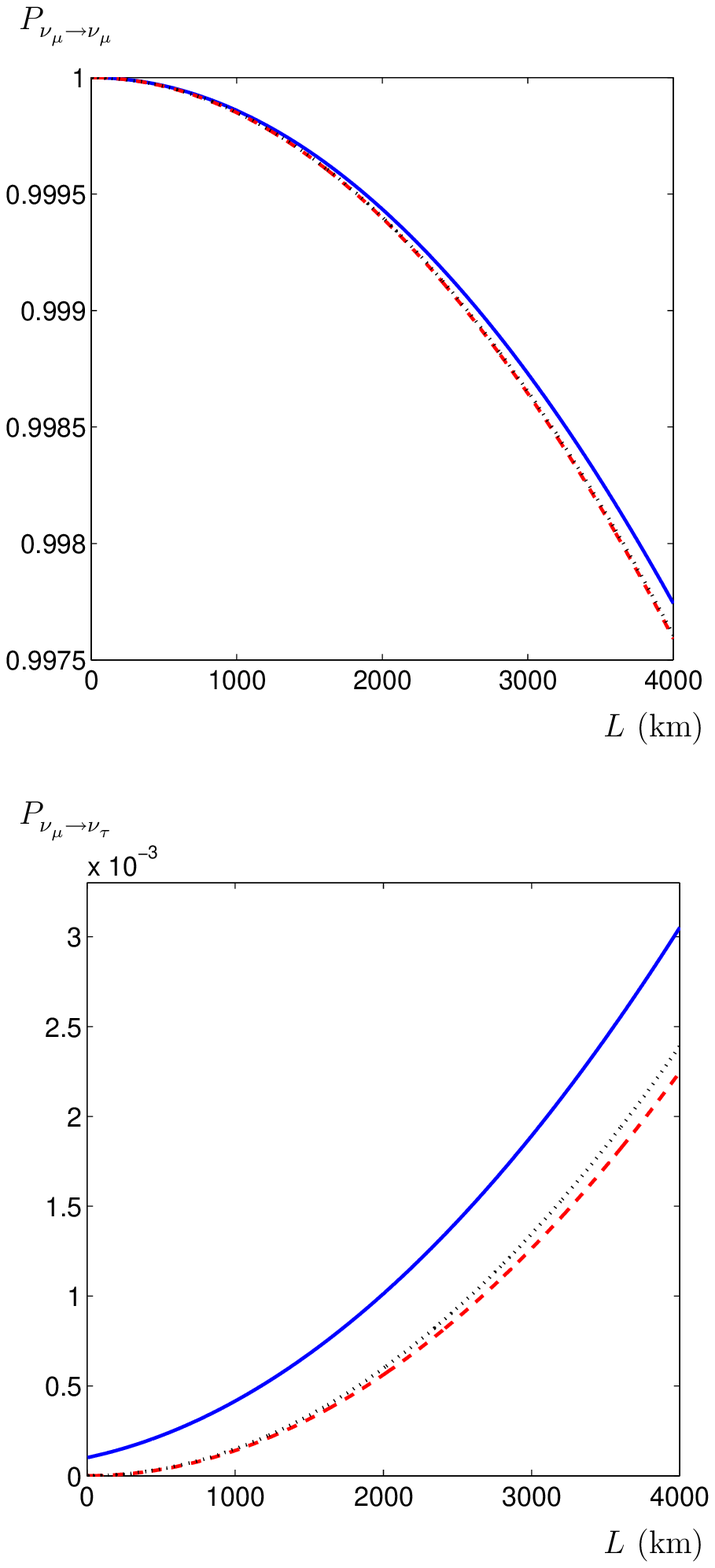}
\vspace{2.4cm}\caption{The probabilities of neutrino oscillation
$\nu_{\mu}^{} \rightarrow \nu_{\mu}^{}$ (the first plot) and
$\nu_{\mu}^{} \rightarrow \nu_{\tau}^{}$ (the second plot) in matter
as functions of the baseline $L$ in Case I (the unitary case,
represented by dashed lines) and Case II (the non-unitary case,
represented by solid lines), in the normal hierarchy case. Here we
choose $E_{\nu}^{} = 50 ~ {\rm GeV}$. The dotted lines in the figure
show the corresponding probabilities if the neutrino mixing is the
exact tri-bimaximal mixing.}
\end{center}
\end{figure}

\begin{figure}[h]
\begin{center}
\vspace{8cm} \subfigure[]{
\includegraphics[bbllx=6.5cm, bblly=6.0cm, bburx=15.0cm, bbury=14.2cm,%
width=7.2cm, height=7.2cm, angle=0, clip=0]{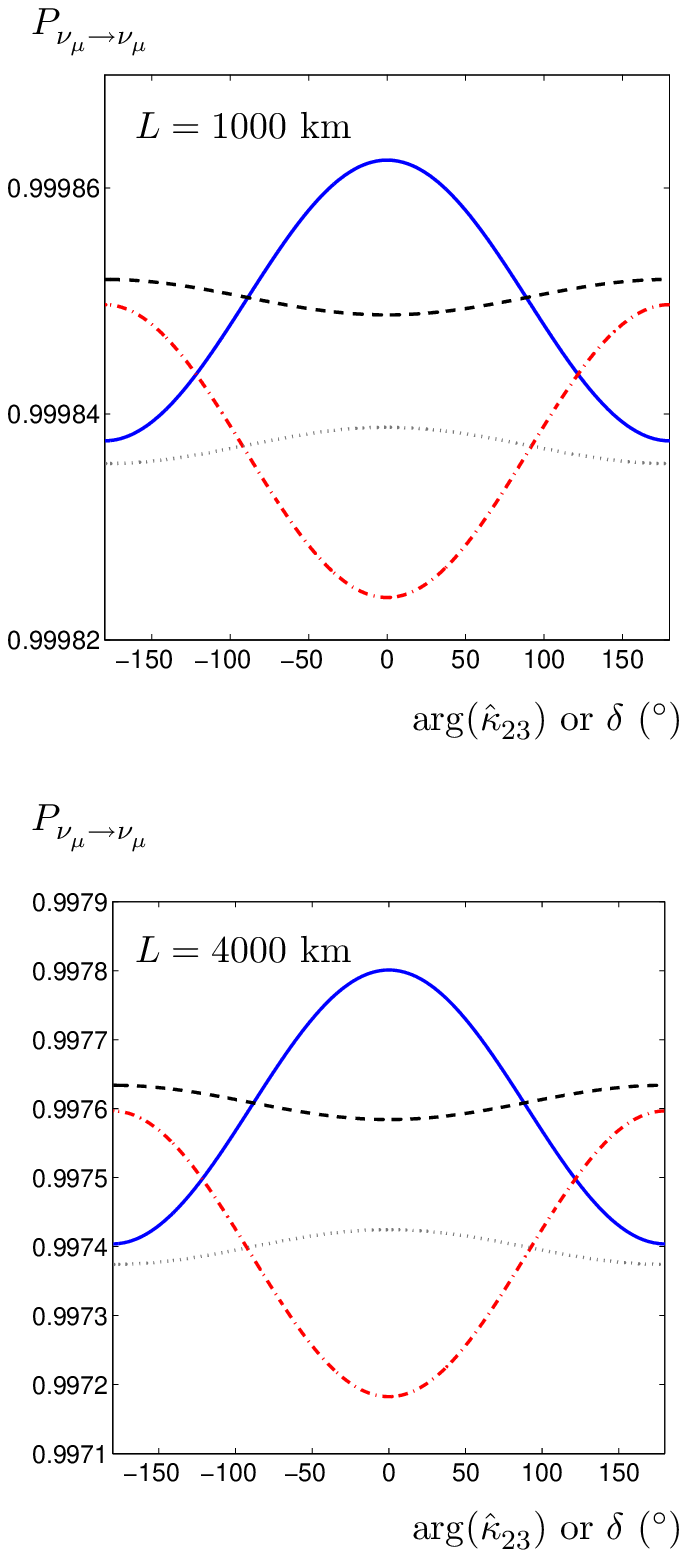}}
\subfigure[]{\includegraphics[bbllx=6.5cm, bblly=6.0cm, bburx=15.0cm, bbury=14.2cm,%
width=7.2cm, height=7.2cm, angle=0, clip=0]{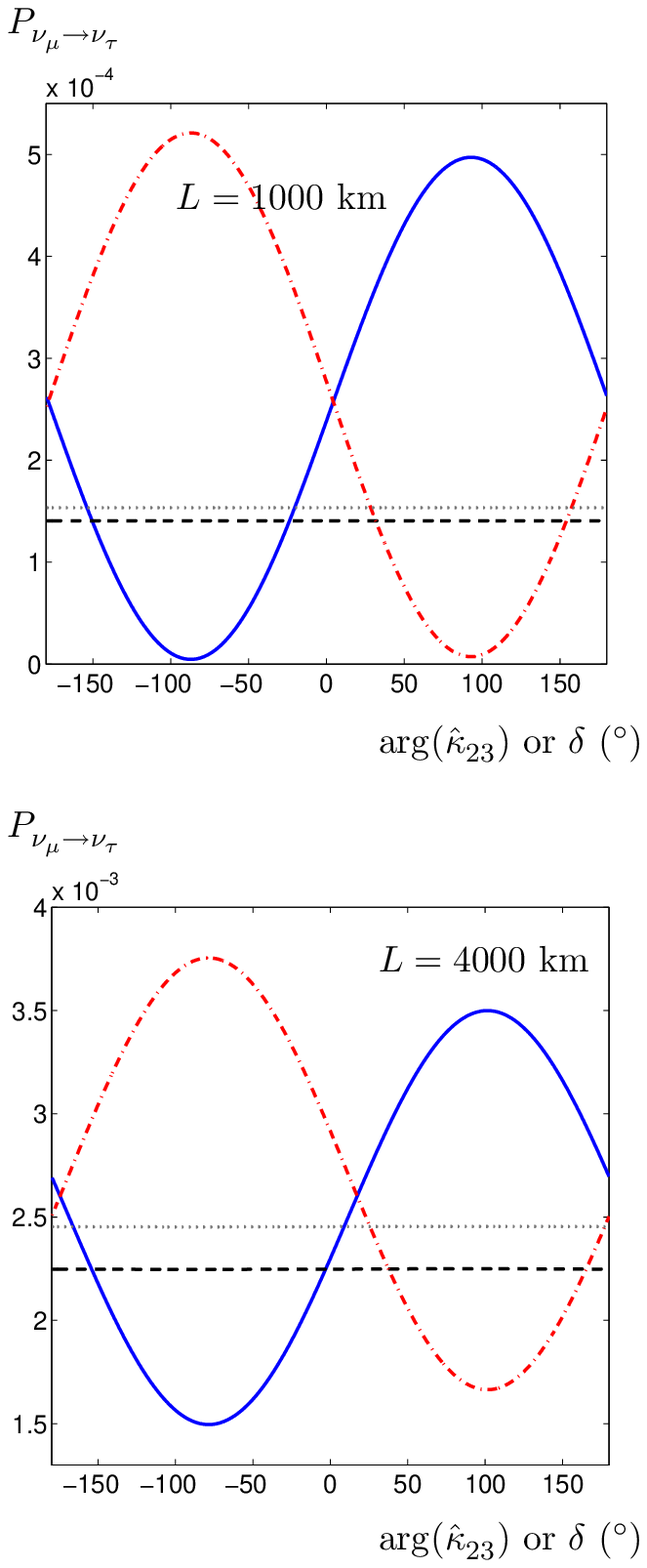}}
\vspace{0.6cm}\caption{The probabilities of neutrino oscillation
$\nu_{\mu}^{} \rightarrow \nu_{\mu}^{}$ (a) and $\nu_{\mu}^{}
\rightarrow \nu_{\tau}^{}$ (b) in matter as functions of the Dirac
phase $\delta$ in Case I (the unitary case, the dashed lines for the
normal hierarchy, the dotted lines for the inverted hierarchy) or
the phase of $\hat{\kappa}_{23}^{}$ in Case II (the non-unitary
case, the solid lines for the normal hierarchy, the dot-and-dash
line for the inverted hierarchy), where we choose $E_{\nu}^{} = 50 ~
{\rm GeV}$.}
\end{center}
\end{figure}

\end{document}